\def\beq{\begin{equation}}
\def\eeq{\end{equation}}
\def\bea{\begin{eqnarray}}
\def\eea{\end{eqnarray}}
\def\bq{\begin{quote}}
\def\eq{\end{quote}}
\def \lsim{\mathrel{\vcenter
     {\hbox{$<$}\nointerlineskip\hbox{$\sim$}}}}

\def\gappeq{\mathrel{\rlap {\raise.5ex\hbox{$>$}}
{\lower.5ex\hbox{$\sim$}}}}
\def\lappeq{\mathrel{\rlap{\raise.5ex\hbox{$<$}}
{\lower.5ex\hbox{$\sim$}}}}

\def\snu{\tilde{\nu}}
\def\snuvi{\langle\tilde{\nu}_i\rangle}

\def\Rpv{R_p \! \! \! \! \! \! /~~}
\def\mnu{[m_{\nu}]_{ij}}
\def\LI{$\{ L_I \} ~$}

\documentstyle [12pt,epsf,epsfig,axodraw]{article}
\begin{document}
\thispagestyle{empty}
\begin{flushright}
{OUTP-00-45P}
\end{flushright}

\vspace{1cm}
\begin{center}
{\large { \bf Basis Independent Neutrino 
   masses in the $R_p$ violating MSSM } }
\end{center}
\vspace{1cm}

\begin{center}
{\bf Sacha Davidson$^{(1)}$, Marta Losada$^{(2)} $ \\}

\vspace{.3cm}
{$^{(1)}${\it 
Theoretical Physics,  Oxford
University, 1 Keble Road, Oxford, OX1 3NP, United Kingdom}}
\vspace{.3cm}
{$^{(2)} $ {\it
 Centro de Investigaciones, 
Universidad Antonio Nari\~{n}o, Cll. 59 No. 37-71, Santa Fe de Bogot\'{a},
Colombia }}
\end{center}

\begin{abstract} 
We  calculate the neutrino mass matrix  up to one loop order
in the MSSM without  $R$-parity,
including the bilinears in the mass insertion approximation.
This introduces additional diagrams usually neglected in
the literature. We systematically  consider the 
possible new diagrams, and find a few 
missing from  our previous work. We provide analytic expressions
for the mass matrix elements in the neutrino flavour
basis, which are independent of the $H_d - L_i$
basis choice in the Lagrangian. We compare the contributions
from different diagrams, and make ``Lagrangian-basis-independent''
estimates of when the new diagrams
need to be included. We briefly discuss
the phenomenology of a toy model of bilinear R-parity violation. 
\end{abstract}


\section{Introduction}

The construction of a model which generates 
neutrino masses \cite{neutrinos} requires some extension
of the Standard Model. Supersymmetry \cite{SUSY,H+K} is 
an interesting possibility  which  provides
new (s)particles. 
We consider the Minimal Supersymmetric Standard Model (MSSM)
without imposing $R$-parity. The quantum number $R_p$ is defined as 
$R_p=(-1)^{L+3B+2S}$, where $L$, $B$, $S$ are the
  lepton and
 baryon number and the spin of the particle, 
respectively \cite{fayet}. In a general $R_p$
non-conserving supersymmetric 
version of the SM, lepton number
is not conserved\footnote{Baryon number is 
also in general not conserved,
however we will impose baryon number conservation 
by hand to avoid proton decay constraints, 
see {\it e.g.},  \cite{dreiner}},  which allows
non-zero neutrino masses.

 The main physical
motivation for neutrino masses comes from the anomalous 
neutrino data from both
solar and atmospheric neutrino 
experiments \cite{ska}. In the R-parity
violating ($\Rpv$)  MSSM,
one neutrino can acquire a tree level mass through
a see-saw effect from the mass matrix of the neutrinos 
and neutralinos \cite{Nilles}.
In order that the rest of the neutrinos become 
massive,  loop diagrams that violate lepton number 
(by two units) must also
be considered \cite{hall,GH}.    
For the model to generate neutrino
masses that fit experimental constraints 
we must require that: $\Delta m^2_{{\mathrm{sun}}} \lsim 10^{-4}$
 eV$^2$, 
 $\Delta m^2_{{\mathrm{atm}}} \in [10^{-3},10^{-2}]$ eV$^2$. 
For solar neutrino data there are several different possibilities for the relevant mixing angle (large or small
mixing angle solutions) while for atmospheric neutrinos the mixing angle must be in
the range $\sin^2 2\theta_{{\mathrm{atm}}} \in [0.85,1]$.

  In the MSSM, the down-type
Higgs and  the lepton
doublet superfields have the same gauge quantum numbers, which means that they
 can   mix if lepton number is not conserved.
Thus, we can construct a vector $L_J = (H_d, L_i)$ 
with $J:4..1 $.
With this notation, 
the superpotential for the  supersymmetric
SM with   $R_p$ violation  can be written as
\beq
W= \mu^J {H}_u  L_J + \frac{\lambda^{JK \ell}}{2} L_J L_K E^c_{\ell} + 
\lambda^{'Jpq} L_JQ_p D^c_q  + h_t^{pq} {H}_u Q_p U^c_q \label{S} .
\eeq
We contract SU(2) doublets with $\varepsilon_{\alpha \beta}$: $
\varepsilon_{11} = \varepsilon_{22} = 0$, $
\varepsilon_{12} = -\varepsilon_{21} = -1$.
The $R_p$ violating and
conserving coupling constants
have been assembled into vectors and matrices in $L_J$ space:
we call the usual $\mu$ parameter $\mu_4$,
and identify  
$h_e^{jk} = \lambda^{4jk}$,\footnote{
 We have changed convention with
respect to 
factors of 2 in $\lambda$ from our previous paper: we now put
$\lambda/2$ in the superpotential, with $\lambda^{4ij} =  h_e^{ij}$.}
  and
$h_d^{pq} = \lambda^{' 4pq}$.
We call the usual $\Rpv$ mass $\epsilon_i =  \mu_i$.
Lower case roman indices $i,j,k$ and $p,q$  are lepton and
quark generation indices. We frequently suppress the
capitalised indices, writing
$\vec{\mu} = ( \mu_4, \mu_3, \mu_2, \mu_1)$.
We also include possible $R_p$ violating couplings among
the soft SUSY breaking parameters, which can
be written as
\begin{eqnarray}
V_{soft} & =& \frac{\tilde{m}_u^2}{2} H_u^{\dagger} H_u + \frac{1}{2}
 L^{J \dagger} [\tilde{m}^2_L]_{JK} L^K  + B^J H_u L_J  \nonumber \\
& & ~~ + A^{ups} H_u Q_p U^c_s + 
     A^{Jps} L_J Q_p D^c_s +
    \frac{A^{JKl}}{2} L_J L_K E^c_l + h.c.~~~.  \label{soft}
\end{eqnarray}
Note that we have absorbed the superpotential parameters
 into the $A$ and $B$ terms; $e.g.$ we write $B^4 H_uH_d$
not $B^4 \mu^4 H_u H_d$.
We abusively use capitals for superfields (as in eq. (\ref{S})) and
for their scalar components.

In this notation where the Higgs joins the leptons
as the fourth direction in \LI space, the relative magnitude
of $R_p$ violating and conserving coupling constants
vary with the choice of the Higgs direction.
$R_p$ violation can be understood geometrically:
if all interactions in the
Lagrangian agree which direction is the Higgs ({\it e.g.}
$\hat{H} = \hat{L}_4$), then $R_p$ is a good symmetry.
But if there is misalignment
in \LI space between different couplings
constants (for instance $\lambda^{'ipq} \neq 0$ in
the basis where $\mu_1 =\mu_2 =\mu_3 = 0$)
then $R_p$ is not conserved.
This $\Rpv$ can be parametrised
by basis independent combinations
of coupling constants
\cite{Nilles,sacha,Fer,GH3}.
These invariant
measures of $R_p$ violation in the Lagrangian
are analogous  to Jarlskog invariants  
which parametrise  CP violation.

Early work on $\Rpv$  neutrino masses \cite{hall}
used models where the bilinear $\Rpv$  could
be rotated 
out of the mass terms ($B_i, \mu_i, \tilde{m}^2_{4i},$
refered to as ``bilinears'') 
into the trilinear interactions.  Most
subsequent analytic estimates have followed
the calculation of \cite{hall} and neglected
the $\Rpv$ bilinear masses in the loop
contributions to the neutrino masses. This
does not generate the  complete set of 
one loop diagrams contributing to $\mnu$,
so it is formally inconsistent.
 It would 
be acceptable if neglecting the bilinear contributions
corresponded to neglecting loops suppressed by
some additional small parameter.
However, the size of the bilinears
is basis dependent, so what is neglected depends
on the basis choice, and this basis dependence
of neutrino masses has caused some confusion in the literature.

In a previous paper \cite{DL1}, we systematically analysed
$\Delta L =2$  loop 
contributions to neutrino masses
in the mass insertion approximation \cite{massins}.
 This led us to identify  new diagrams
which have not been included in many analytic
estimates in the literature.
We calculated the neutral loop
diagram exactly (Grossman-Haber diagram),
and gave basis-independent estimates for 
all diagrams using a common SUSY scale
for all R-parity conserving mass parameters.
The mass insertion approximation is
valid as we know  experimentally that neutrino
masses are small.
It is 
a particularly
transparent way to include the  bilinears, because
 the $\Rpv$ Feynman rules can be
read off the Lagrangian, and all the $\Rpv$
appears perturbatively in the numerator (``upstairs'')
 of each contribution. The relative contributions
of different diagrams are easy to see.
The main point of our previous paper was that
the issue of the basis choice for calculating neutrino masses is 
transparent  if we
 keep all contributions.

The bilinear $\Rpv$ contribution to loops, as well
as at tree level, has been included in \cite{num,chunetal,hirschetal},
 where neutrino masses are obtained,  
after the exact diagonalization
of the mass matrices of all particles which 
propagate
in the $\Delta L=2$ loops. 
The bilinears mix Standard Model and SUSY
particles,  generating  mass matrices 
of large dimension, so these results
are partially numerical. They propagate
tree-level mass eigenstates in their loop
diagrams. The discussion in ref. \cite{hirschetal} 
includes gauge loops. Reference \cite{chunetal}
have approximate diagonalisation formulae, calculated
in the seesaw approximation, which is equivalent
to the mass insertion approximation used here; they
have few diagrams  and
long mixing angle formulae, whereas we have
more diagrams
and only MSSM mixing angles.

The purpose of this paper is to provide basis-independent
neutrino mass matrix elements, they  are given
in Appendices A and B.
We present complete analytic expressions
for all of the diagrams, and clarify how
the neutrino mass matrix elements can be
``basis-independent''.  
 We  discuss in detail when the different contributions
to the neutrino mass matrix are relevant, identifying for 
both bilinear and trilinear contributions when they
must be included or can be neglected.
 We discuss phenomenological issues 
for a toy model where the  $R_p$ violation comes from the soft
 SUSY breaking sector of the potential.
We will discuss phenomenology in
more detail in a subsequent publication \cite{AMS}.
In the literature, most analyses using neutrino
 data have taken into account only
the effects of $R_P$ violation from trilinears
 and the $\mu$ bilinear \cite{everyone}.

In section \ref{sect2}, we discuss what we mean by ``basis independent'' 
neutrino mass matrix elements, what we calculate, and how basis dependent
it is. In section \ref{sect3}, we discuss when the bilinears
are important in the loop contributions to
neutrino masses. This is a detailed discussion,
with generation indices, of results presented in
\cite{DL1}. In section \ref{pheno},
we study the phenomenology of some
simple bilinear models.
We conclude in section
\ref{concl}. Exact one-loop
neutrino mass matrix elements
are presented in Appendix A  in the $\snuvi = 0$
basis. In Appendix B, the neutrino mass matrix elements
are given in terms of MSSM parameters and basis-independent 
combinations of coupling constants 
that parametrise $\Rpv$.
 Additional Feynman rules for including $\Rpv$ in the
mass insertion approximation can be found in Appendix
\ref{C1}. Numerical bounds on R-parity violating parameters 
obtained from  the different
diagrams are given in Appendix \ref{D1}.

\section{Issues of basis}
\label{sect2}

There is no unique interaction eigenstate basis 
 for the $Y = -1$ fields (the components of
the $L_i$ and $H_d$ superfields) in the $\Rpv$  MSSM.  This means that
the Lagrangian parameters depend on
an arbitrary choice of basis, or equivalently that
Lagrangians which differ by
a rotation in $\{ L_I \}$ space
make equivalent physical predictions.
Observables cannot depend on the choice
of basis in the Lagrangian $\cal L$, so they must be
scalar functions
of the vector and tensor parameters in  $\{ L_I \}$
space.
It is common in SUSY to
study the dependence
of physical observables on
 Lagrangian inputs. The basis-dependence of
the Lagrangian makes this
problematic in $\Rpv$ models
because a point in physical parameter space corresponds
to different numerical inputs in different
bases. So comparing results calculated in
different bases is difficult.

The $\Rpv$ in the Lagrangian can be
parametrised in a basis independent way
by combining coupling constants, masses and
vacuum expectation values (vectors and matrices
in  $\{ L_I \}$ space) into
scalar ``invariants'', which are zero when
$R_p$ is conserved.  Physical observables can be expressed
in terms of these invariants and other $R_p$ conserving
quantities.  For a discussion of constructing
invariants, see {\it e.g.} \cite{sacha,DL1,GH3,DLR}. 

Neutrino masses and mixing angles
are measurable quantities, so must be (Lagrangian) 
basis-{\em independent}. The neutrino mass matrix
$\mnu$, $i,j:1..3$, is basis {\em dependent},
depending on
the choice of Higgs direction in
\LI space, and on the choice of lepton flavour directions
$\hat{L}_i$ and $\hat{L}_j$.  A physically motivated basis choice
for the $\hat{L}_i$ is the charged lepton mass eigenstate basis
($\equiv$ the neutrino flavour basis), because the neutrino
masses and mixing angles can
be computed from $\mnu$ in this basis, so it contains all
the measurable information. We can express
the charged lepton mass
eigenstate vectors in a basis-independent
way, using the mass insertion approximation, so
we present {\it Lagrangian }
basis  independent expressions for the elements of
$\mnu$ in the flavour basis.
 So we have  ``basis-independent''
expressions for matrix elements which are
intrinsically basis-dependent---this is sensible
because we want expressions which are
independent of the arbitrary Lagrangian basis
choice; the flavour basis is ``physical''.

The basis we work in is
\bea
\hat{H}_d = \frac{\vec{v}}{v}    \nonumber \\
\hat{L}^i = \frac{\lambda^i \cdot \vec{v} }{|\lambda^i \cdot \vec{v}|},
\label{basis}
\eea
 where $\vec{v}$ is the vector of  
vacuum expectation values in the down-type Higgs sector and we
impose the requirement that $ \vec{v} \cdot \lambda^i \cdot 
\lambda^j \cdot \vec{v} \propto \delta^{ij}$.
The lower case index labelling the matrix
$\lambda^k$ is the singlet lepton index: $[\lambda^k]_{IJ} = \lambda^{IJk}$.
So there are three flavour eigenstates $\hat{L}^i$ in
\LI space:   $\hat{L}^i_J =    \lambda^{JKi} v_K
/\sqrt{v_P \lambda^{NPi} \lambda^{NMi}v_M}$  (no sum on $i$)
\footnote{The capitalised index ordering on
$\lambda^{NPi}$ is because $\vec{a} \cdot \lambda \cdot \vec{b}$
 $= -\vec{b} \cdot \lambda \cdot \vec{a}$ is an antisymmetric
product. The transpose of $\hat{L}^i$ is $ -\vec{v} \cdot \lambda^i/
|\vec{v} \cdot \lambda^i|$. }.
This corresponds to
the charged lepton mass eigenstate
basis in the absence of $\Rpv$. In this $R_p$ conserving case 
$\lambda^{4ij} =  h_e^{ij}$ and $\lambda^{kij} = 0,$ where $
k,i,j:1..3$. So $  \lambda^{IJk}v_I = \sqrt{2} m_e^{jk}$
($I = 4, J = j:1..3$), $\vec{v} \cdot \lambda^i \cdot \lambda^j \cdot
\vec{v} \propto \delta^{ij}$ is the condition
that the $\{e_R^c\}$ are in the mass
eigenstate basis, and $\lambda^i \cdot \vec{v}  $ are the
charged doublet eigenvectors. 
 In
the presence of bilinear $\Rpv$, the basis (\ref{basis})
will not be exactly
the charged lepton mass eigenstate basis. However, bilinear $\Rpv$ masses
are required to be small by neutrino masses, so
the basis (\ref{basis}) is close to the
charged lepton mass eigenstate basis. We include
the correction due to the bilinears in the mass insertion 
approximation.

Neutrino masses are proportional to $\Rpv$ couplings,
and observed to be small. We
can therefore compute neutrino masses by
perturbing in $\Rpv$  masses and trilinear couplings. 
We include the bilinear $\Rpv$
masses in the mass insertion approximation, which
means  all $\Rpv$ couplings appear in the numerator (``upstairs'')
of the expression corresponding to a diagram.
We propagate MSSM mass eigenstates, whose masses
we make ``basis-independent''  by
substituting $ B_4 \rightarrow  \vec{B} \cdot \vec{v} /|\vec{v}|,$
$ \mu_4 \rightarrow  \vec{\mu} \cdot \vec{v} /|\vec{v}|,$ and 
$\tilde{m}^2_{44}  \rightarrow  \vec{v} \cdot \tilde{m}^2
 \cdot \vec{v} /|\vec{v}|^2$.
The physically relevant flavour basis is written
in terms of vevs and coupling constants (see equation
\ref{basis}), so neutrino mass matrix elements in
this basis are proportional to invariants. 
We list the four  invariants relevant for neutrino
masses
in table \ref{t2}, along with their value
in the basis where the sneutrinos have no vev.

\begin{table}
$
\setlength{\arraycolsep}{2em}
\begin{array}{||c|c||}
\hline
\hline
 & \\
\delta_{\mu}^i  \equiv \frac{ \vec{\mu} \cdot  \lambda^i \cdot  \vec{ v}}{|  \vec{\mu}| \sqrt{2} m_e^i} 
&
\frac{\mu^i}{|\mu|} \\
 & \\
\hline
 &  \\
\delta_{B}^i \equiv \frac{  \vec{B} \cdot  \lambda^i  
  \cdot  \vec{v}}{|  \vec{B}|  \sqrt{2} m_e^i} & \frac{B^i}{|B|}\\
 &  \\
\hline
 &  \\
\delta_{\lambda'}^{ipq} \equiv \frac{  \vec{\lambda}^{'pq}  \cdot 
\lambda^i  \cdot   \vec{v}}
{ \sqrt{2} m_e^i} & {\lambda}^{'ipq}\\
 &  \\
\hline
 &  \\
\delta_{\lambda}^{ijk} \equiv \frac{ \vec{v}  \cdot \lambda^{i}
\lambda^{k} \lambda^j  \cdot  \vec{v}}
{2 m_e^i m_e^j} &{\lambda}^{ijk} \\
 &  \\
\hline
\hline
\end{array}
$
\caption{The basis-independent invariants used
to parametrise the $\Rpv$ relevant for
neutrino masses, together with
their value in the $\snuvi = 0$ basis. 
They are zero if $R_p$ is conserved.
$\delta_{\mu}$ and $\delta_{B}$ parametrise
bilinear $\Rpv$. Note
that these invariants have signs: for arbitrary vectors
 $ \vec{a}$ and   $ \vec{b}$,
  $ \vec{a} \cdot \lambda^{i}  \cdot \vec{b} = -
  \vec{b}  \cdot  \lambda^{i}  \cdot \vec{a} $.}
\label{t1}
\end{table}

A significant advantage of the basis-independent
formalism is that the neutrino masses can
be computed in any Lagrangian basis, including
those where the mass insertion approximation is not valid. 
This could be easier than rotating to
the $\snuvi = 0$ basis, because the vev is a derived
quantity, calculated by minimising
the potential. For instance,  consider
a model where
all the $\Rpv$ is in $\vec{B}$ and
$\vec{v}$---that is, there is a basis
where all the $\Rpv$ couplings
other than $B_i$ and $\snuvi$ are zero. 
Clearly it is easier to
evaluate invariants in this
basis than to rotate to $\snuvi = 0$.
The basis independent formalism also makes
it simple to compare results computed 
in different bases.

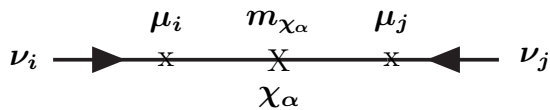
\begin{figure}[htb]
\unitlength1mm
\SetScale{2.8}
\begin{boldmath}
\begin{center}
\begin{picture}(60,20)(0,0)
\ArrowLine(0,0)(15,0)
\Line(45,0)(15,0)
\ArrowLine(60,0)(45,0)
\Text(-2,0)[r]{$\nu_i$}
\Text(62,0)[l]{$\nu_j$}
\Text(30,-5)[c]{$\chi_{\alpha}$}
\Text(30,5)[c]{$m_{\chi_{\alpha}}$}
\Text(30,0)[c]{X}
\Text(15,0)[c]{x}
\Text(45,0)[c]{x}
\Text(15,5)[c]{$\mu_i$}
\Text(45,5)[c]{$\mu_j$}
\end{picture}
\end{center}
\end{boldmath}
\caption{Tree-level neutrino mass in the mass
insertion approximation}
\protect\label{ft}
\end{figure}

The tree level neutrino mass is non-zero
if $\delta_{\mu} \neq 0$ \cite{Nilles}. The diagram
appears in figure \ref{ft} in the
mass insertion approximation.  In the basis (\ref{basis}),
it contributes a mass matrix
\beq
\mnu ^{tree}  =  - ( \vec{\mu} \cdot \hat{L}_i) 
 \sum_{\alpha}  \frac{ Z^*_{\alpha 3}Z^*_{ \alpha 3} }
  {m_{\chi_{\alpha}}} (\vec{\mu} \cdot \hat{L}_j),
\label{mtree}
\eeq
which gives a mass
$m_3^{tree} = \sum_{i, \alpha} (\delta_{\mu}^i)^2 
Z_{ \alpha 3}^{*2} |\mu|^2/m_{\chi_{\alpha}}$ to the neutrino
\beq
\hat{\nu}_3^{tree} = \frac{\delta_{\mu}^i}{\delta_{\mu}} \hat{L}_i,
\label{nu3}
\eeq
where $\delta_{\mu} = \sqrt{\sum_i (\delta_{\mu}^i)^2},$
and the neutralino index $\alpha$ runs from 1 to 4.
The index ``3'' on $Z$ corresponds to the interaction
eigenstate $\tilde{h}_u$---see Appendix C for
our conventions on $Z$. Equation (\ref{mtree}) is equivalent
to the usual formula $ m_{\nu}^{tree} = det[M^{(5)}]/det[M^{(4)}]$,
where $M^{(5)}$ ($M^{(4)}$) is the $\Rpv$ $ 5 \times 5$ neutralino
and neutrino mass matrix (MSSM neutralino mass matrix), as
can be seen by writing $M^{(5)}$ and $M^{(4)}$ in the MSSM
mass eigenstate basis with $\snuvi = 0$.

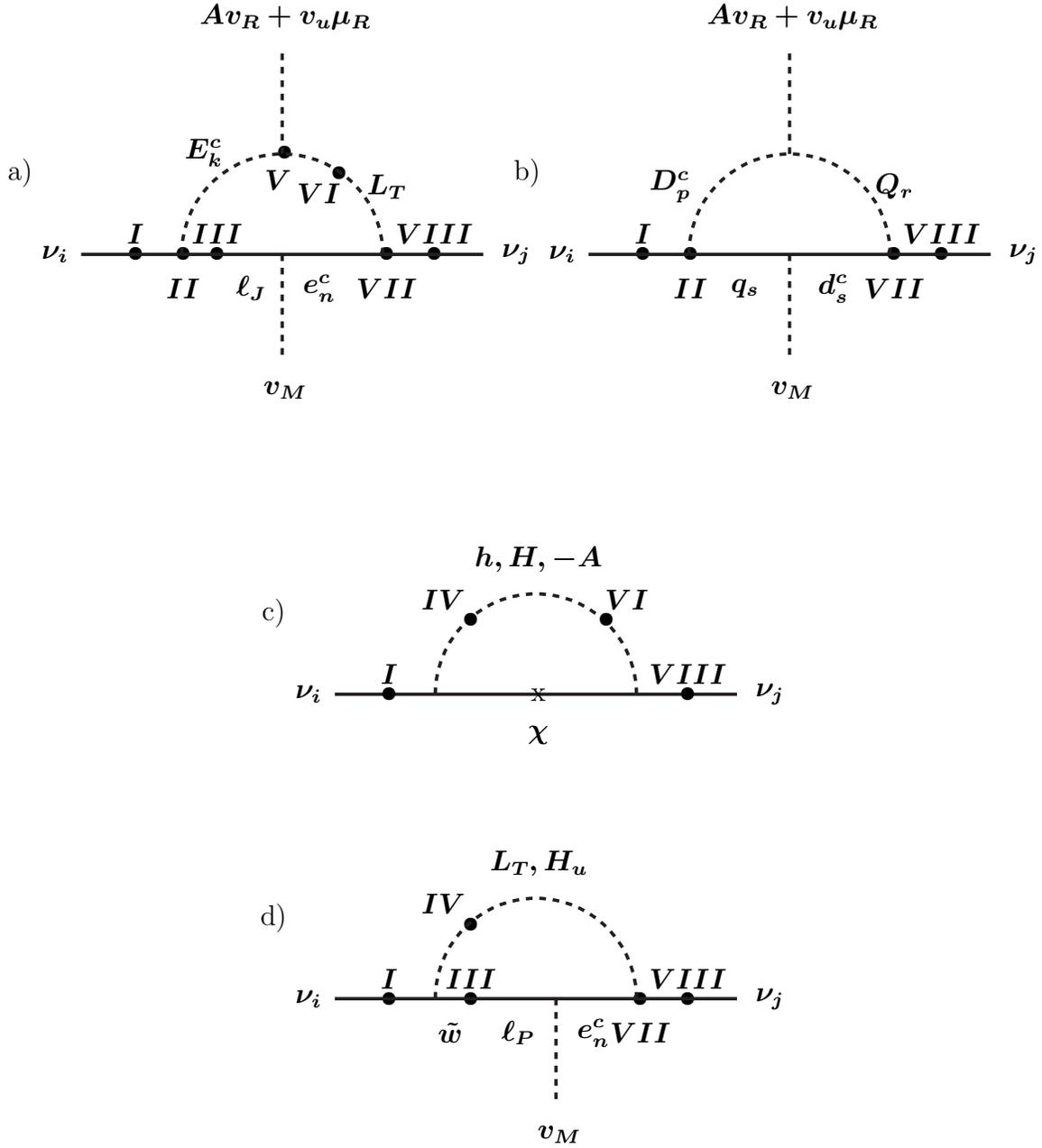
\begin{figure}[htb]
\unitlength1mm
\SetScale{2.8}
\begin{boldmath}
\begin{center}
\begin{picture}(60,60)(0,-20)
\Text(-9,12)[c]{a)}
\Text(8,0)[c]{$\bullet$}
\Text(8,3)[c]{$I$}
\Text(15,0)[c]{$\bullet$}
\Text(15,-5)[c]{$II$}
\Text(20,0)[c]{$\bullet$}
\Text(20,3)[c]{$III$}
\Text(45,0)[c]{$\bullet$}
\Text(45,-5)[c]{$VII$}
\Text(30,15)[c]{$\bullet$}
\Text(29,11)[c]{$V$}
\Text(52,0)[c]{$\bullet$}
\Text(52,3)[c]{$VIII$}
\Text(38,12)[c]{$\bullet$}
\Text(35,9)[c]{$VI$}
\Line(0,0)(15,0)
\Line(45,0)(15,0)
\Line(60,0)(45,0)
\DashCArc(30,0)(15,0,180){1}
\Text(-2,0)[r]{$\nu_i$}
\Text(62,0)[l]{$\nu_j$}
\Text(18,15)[c]{$E^c_k$}
\Text(45,10)[c]{$L_T$}
\Text(25,-5)[c]{$\ell_J$}
\Text(35,-5)[c]{$e^c_n$}
\Text(30,-20)[c]{$v_M$}
\DashLine(30,0)(30,-15){1}
\Text(30,35)[c]{$A v_R +  v_u \mu_R$}
\DashLine(30,15)(30,30){1}
\end{picture}
\hspace{12mm}
%
\begin{picture}(60,60)(0,-20)
\Text(-9,12)[c]{b)}
\Text(8,0)[c]{$\bullet$}
\Text(8,3)[c]{$I$}
\Text(15,0)[c]{$\bullet$}
\Text(15,-5)[c]{$II$}
\Text(45,0)[c]{$\bullet$}
\Text(45,-5)[c]{$VII$}
\Text(52,0)[c]{$\bullet$}
\Text(52,3)[c]{$VIII$}
\Line(0,0)(15,0)
\Line(45,0)(15,0)
\Line(60,0)(45,0)
\DashCArc(30,0)(15,0,180){1}
\Text(-2,0)[r]{$\nu_i$}
\Text(62,0)[l]{$\nu_j$}
\Text(12,10)[c]{$D^c_p$}
\Text(45,10)[c]{$Q_r$}
\Text(23,-5)[c]{$q_s$}
\Text(36,-5)[c]{$d^c_s$}
\Text(30,-20)[c]{$v_M$}
\DashLine(30,0)(30,-15){1}
\Text(30,35)[c]{$A v_R +  v_u \mu_R$}
\DashLine(30,15)(30,30){1}
\end{picture}
\end{center}
%
\begin{center}
\begin{picture}(60,40)(0,0)
\Text(-9,12)[c]{c)}
\Line(0,0)(15,0)
\Line(45,0)(15,0)
\Line(60,0)(45,0)
\DashCArc(30,0)(15,0,180){1}
\Text(8,0)[c]{$\bullet$}
\Text(8,3)[c]{$I$}
\Text(20,11)[c]{$\bullet$}
\Text(16,14)[c]{$IV$}
\Text(52,0)[c]{$\bullet$}
\Text(52,3)[c]{$VIII$}
\Text(40,11)[c]{$\bullet$}
\Text(43,14)[c]{$VI$}
\Text(-2,0)[r]{$\nu_i$}
\Text(62,0)[l]{$\nu_j$}
\Text(30,20)[c]{$h,H,-A$}
\Text(30,0)[c]{x}
\Text(30,-6)[c]{$\chi$}
\end{picture}
\end{center}
\end{boldmath}
\begin{boldmath}
\begin{center}
\begin{picture}(60,60)(0,-20)
\Text(8,0)[c]{$\bullet$}
\Text(8,3)[c]{$I$}
\Text(20,0)[c]{$\bullet$}
\Text(20,3)[c]{$III$}
\Text(45,0)[c]{$\bullet$}
\Text(45,-5)[c]{$VII$}
\Text(52,0)[c]{$\bullet$}
\Text(52,3)[c]{$VIII$}
\Text(20,11)[c]{$\bullet$}
\Text(16,14)[c]{$IV$}
\Line(0,0)(15,0)
\Line(45,0)(15,0)
\Line(60,0)(45,0)
\DashCArc(30,0)(15,0,180){1}
\Text(-9,12)[c]{d)}
\Text(-2,0)[r]{$\nu_i$}
\Text(62,0)[l]{$\nu_j$}
\Text(30,20)[c]{$L_T, H_u$}
\Text(17,-5)[c]{$\tilde{w}$}
\Text(27,-5)[c]{$\ell_P$}
\Text(38,-5)[c]{$e^c_n$}
\Text(33,-20)[c]{$v_M$}
\DashLine(33,0)(33,-15){1}
\end{picture}
\end{center}
\end{boldmath}
\caption{Schematic 
representation of one-loop diagrams
contributing to neutrino masses,
in a Lagrangian basis. The blobs
indicate possible positions for $\Rpv$ interactions, which
can be 
trilinears (at positions II and VII) or mass insertions.
The misalignment between $\vec{\mu}$ and
$\vec{v}$ allows a mass insertion on
the lepton/higgsino lines (at points
I, III, or VIII) and at the $A$-term
on the scalar line (position V). The
soft $\Rpv$ masses appear as mass insertions
at positions VI and IV on
the scalar line. 
Figure a) is the charged loop with trilinear couplings
$\lambda$ (or $h_e$)  at the vertices.
Figure b)  is the  coloured loop with trilinear $\lambda'$ or
yukawa $h_b$  couplings.
Figure c) is the neutral  loop 
with two gauge couplings (in the MSSM
mass eigenstate basis), and 
figure d) is the charged loop
 with one gauge and and  a Yukawa coupling. This diagram occurs
if gauginos mix with  charged leptons---that is if
$\delta_{\mu} \neq 0$.} 
\label{fl}
\end{figure}

 Loop corrections to the neutrino
mass matrix can be divided into three
categories. First, there are gauge and top Yukawa coupling loops
which renormalise the mass of $\nu_3^{tree}$---
we  neglect these
because their effect is small \cite{num}.
Secondly, there are  loop corrections to $\delta_{\mu}^i$
(equivalently, $\mu_i$ and/or $ \snuvi$)
which can modify the direction of $\hat{\nu}_3$---
these we partially compute and list in the Appendices.
They are renormalisation scale dependent,
because they are loop corrections to
the tree mass. We are not interested in
loop corrections to the tree mass, so we
rotate these away. This is discussed in more
detail in Appendix \ref{loop}.
Finally, there are finite loops which
give mass to any neutrino. The third group  
are the  most interesting loops,
because they generate mass for
the  two neutrinos who are massless at tree level.
Schematic representations of the one-loop
neutrino mass diagrams that we consider
are reproduced in figure \ref{fl}. This is
a slightly modified version of a figure from ref. \cite{DL1}.
Each of the four diagrams
represents a number of
Feynman diagrams. This should be an almost\footnote{ 
We discuss in Appendix \ref{B1} why we neglect certain finite
diagrams of the third type.}
complete set of
one loop, $\Delta L = 2$ diagrams 
which generate mass matrix
elements for the  neutrinos that are massless at
tree level.
The possible places on the diagram
for the two required lepton number violating
interactions are labelled
$ I..VIII$. With the definition
$\hat{H} \propto \vec{v}$,
lepton number violation is not possible
at  the charged lepton mass insertion in
diagrams a) and d). (A sneutrino vev could
provide lepton number violation at this
point in a basis where $\snuvi \neq 0$.)

In Appendix \ref{A1}, we list the loop diagrams, 
giving exact formulae in the $\snuvi = 0$ basis
for each diagram's contribution to $\mnu$.
The basis-independent version of these
formulae is in Appendix \ref{B1}. 
In table \ref{t2}, we summarize the diagrams and make 
basis independent estimates  by setting
all the heavy masses to a unique scale $m_{SUSY}$.
 The lists starts with
the canonical trilinear diagrams, then the Grossman-Haber loop
induced by the soft bilinears, and 
finally all the additional diagrams which arise
when the bilinear $\delta_{\mu}^i$
are included in the loops. For each diagram, labelled by $a \rightarrow d$
for the diagram category of figure \ref{fl}, and
by roman numeral identifying
where the $\Rpv$ should appear on the figure, we give
an estimate of the diagram in terms of the
invariants listed in table \ref{t1}. 
We have added a few
relevant diagrams
missing from the original list. 
In Appendix B, we make an exhaustive list of 
all potential diagrams, explain why
some are zero, and give basis-independent
expressions.

\begin{table}
$
\setlength{\arraycolsep}{1em}
\begin{array}{|c|||c|c|c||}
\hline
\hline
No. &{\rm diagram} &{\rm position~ of} ~ \Rpv &
16 \pi^2 m_{SUSY} [m_v]^{ij} \\ \hline \hline
1&a & II~~~ VII 
 &
  \delta_{\lambda}^{ink} 
 \delta_{\lambda}^{jkn} 
  m_{e_{n}} m_{e_{k}} 
 \\
 \hline
2&b & II ~~~ VII
 & 
 3 \delta_{\lambda'}^{iqq}
 \delta_{\lambda'}^{jqq}
      (m_{d_{q}})^2 
 \\
 \hline
 \hline
 \hline
3 &c & IV ~~~ VI
 &
 g^{2} \delta_B^{i}
 \delta_B^{j} m_{\chi} m_{SUSY}/4
 \\
 \hline \hline
\hline
4 &b&I~~VII + II~~VIII
 &
 3 (\delta_{\mu}^{i}\delta_{\lambda'}^{jqq} + 
   \delta_{\mu}^{j}\delta_{\lambda'}^{iqq}) 
       (m_{d_{q}})^2  h_d^{q} 
 \\
 \hline
5&a& II~~~VI
 &
 \delta_{\lambda}^{ijk}
 m_{e_k} \delta_B^k    (  m_{e_{j}} h_e^{j} -  m_{e_{i}}  h_e^{i})
 \\
 \hline
6 &a&I~~VII + II~~VIII
 &
 (\delta_{\mu}^{i}\delta_{\lambda}^{jkk} + 
 \delta_{\mu}^{j}\delta_{\lambda}^{ikk}) 
      (m_{e_{k}})^2 
  h_e^{k} 
 \\
 \hline
7 &a&I~~~V&
 \delta_{\mu}^{i}\delta_{\mu}^{j} 
       ( (m_{e_{j}} h_e^{j})^2 
 +(m_{e_{i}} h_e^{i} )^2    ) 
 \\
 \hline
8 &a&II~~~V
 & 
   \delta_{\lambda}^{ijk}\delta_{\mu}^k  m_{e_k}   
  (h_e^i  m_{e_i}-h_e^j  m_{e_j})
 \\
 \hline
9 &a& III~~~V
 & 
    \delta_{\mu}^i  \delta_{\mu}^j  
 m_{e_j} m_{e_i}   h_e^{i}  h_e^{j}     
 \\
 \hline
10 &a& III~~~VIII
 &
   \delta_{\mu}^i  \delta_{\mu}^j 
 ( (m_{e_i}   h_e^{i}) ^2 +  (m_{e_j}   h_e^j)^2)      
 \\
 \hline
11 &a& I~~~VI&
   \delta_{\mu}^i  \delta_{B}^j   (m_{e_j}   h_e^{j})^2   
  +  \delta_{\mu}^j  \delta_{B}^i   (m_{e_i}   h_e^{i})^2   
 \\
 \hline
12 &a& III~~~VII&
   \delta_{\lambda}^{jin} \delta_{\mu}^n m_{e_n} 
  (  m_{e_j} 
   h_e^{j} -  m_{e_i} 
   h_e^{i}) 
 \\
 \hline
13 &a& III~~~VI&
   (\delta_B^i  \delta_{\mu}^j h_e^j h_e^i  m_{e_i} m_{e_j} 
 +\delta_B^j  \delta_{\mu}^i h_e^i h_e^j  m_{e_j} m_{e_i})    
 \\
\hline
14 &d& III~~~IV&
 g(  \delta_B^i \delta^{j}_{\mu} (m_{e_j})^2 +
  \delta_B^j \delta^{i}_{\mu} (m_{e_i})^2)
 \\
 \hline
15 &d& III~~~VIII
 &
 g \delta_{\mu}^i  \delta_{\mu}^j
 ( ( m_{e_i})^2 +  ( m_{e_j})^2)
 \\
 \hline
16 &d& I~~~III
 &
 g \delta_{\mu}^i  \delta_{\mu}^j
 ( ( m_{e_i})^2 +  ( m_{e_j})^2)
 \\
 \hline
17 &d& I~~~VII
 &
 g  m_{e_k} m_{SUSY}(\delta_{\mu}^i  \delta_{\lambda}^{jkk}+\delta_{\mu}^j  \delta_{\lambda}^{ikk})
 \\
 & & 
 &
3  g   m_{d_k} m_{SUSY}(\delta_{\mu}^i  \delta_{\lambda'}^{jkk}+\delta_{\mu}^j  \delta_{\lambda'}^{ikk})
 \\
 \hline
18 &d& III~~~VII
 &{\rm zero~for~degenerate~ sleptons}
 \\
 \hline
19 &c&  I~~~VI+IV~~~VIII
 &
 g^2  m_{SUSY}^2(\delta_{B}^i  \delta_{\mu}^{j}+\delta_{\mu}^i  \delta_{B}^{j})/4
 \\
 \hline
20 &d& I~~~V
 &
 g   \delta_{\mu}^i  \delta_{\mu}^{j} ( ( m_{e_i})^2 +  ( m_{e_j})^2)
 \\
\hline
21 &d& I~~~IV
 &
 g   (\delta_{\mu}^i  \delta_{B}^{j} 
( m_{e_j})^2 + \delta_{\mu}^j  \delta_{B}^{i} ( m_{e_i})^2)
 \\
\hline
\hline
\end{array}
$
\caption{ Estimated contributions to $[m_{\nu}]^{ij}$ from
all the  diagrams. In the second two columns
is the label of the diagram of figure 2, and the position
on the diagram of the two $\Delta L = 1$ interactions. 
Column four is the ``basis independent'' estimated 
 contribution to the neutrino mass matrix 
in the  flavour basis.
All indices other than $i$ and $j$ are summed.}
\label{t2}
\end{table}

To evaluate diagram b) with $\Rpv$ at
points $II$ and $VII$ (this is the usual
coloured trilinear diagram), we identify
the external neutrino legs
as $\hat{L}_i$ and  $\hat{L}_j$. At
vertices $II$ and $VII$ sit
couplings $\vec{\lambda}^{'s{\bf p}}$
and  $\vec{\lambda}^{'{\bf r} s}$. (The bold
face indices correspond to the
squark mass eigenstate basis---see
Appendix \ref{C1}.) The diagram
is therefore
\begin{eqnarray}
[m_{\nu} ]_{kj} &  = & -3 \sum_{s,{\bf p},{\bf r}}( i \hat{L}_k \cdot 
\vec{\lambda}^{'s{\bf p}})  
( i \vec{\lambda}^{'{\bf r}s}  \cdot \hat{L}_j)  
 (i |\tilde{A}^{{\bf p}{\bf r}}_d|) (- | m_{d_s}|)  \nonumber \\
& &    ~~~\times \int \frac{d^4 k}{(2 \pi)^4} 
       \frac{i}{k^2 - m_{\tilde{D}^c_{\bf p}}^2}
       \frac{i}{k^2 - m_{\tilde{Q}_{\bf r}}^2}
      \frac{i}{k^2 - m_{d_s}^{2}}
   + (k \leftrightarrow j)  \nonumber \\
 & = &  - \frac{ 3}{16 \pi^2}\sum_{s,{\bf p},{\bf r}} {\lambda^{'ks{\bf p}}
 \lambda^{'j{\bf r}s}}  
 \lambda^{'Mss} \frac{v_M}{\sqrt{2}} \biggl[ (\lambda A)^{'R{\bf r}{\bf p}} 
    \frac{v_R}{\sqrt{2}}\nonumber \\
& +&
  \frac{v_u}{\sqrt{2}} \mu_R  \lambda^{'R{\bf r}{\bf p}}\biggr]
 I( m_{\tilde{Q}_{\bf r}},  m_{\tilde{D}^c_{\bf p}}, m_{d_s}) + (k \leftrightarrow j)
     \nonumber \\
 & = & - \frac{ 3}{16 \pi^2}\sum_{s,{\bf p},{\bf r}}
  \delta_{\lambda'}^{ks{\bf p}} \delta_{\lambda'}^{j{\bf r}s}
    I( m_{\tilde{Q}_{\bf r}},  m_{\tilde{D}^c_{\bf p}},m_{d_s}) 
|\tilde{A}_d^{{\bf p}{\bf r}}| |m_{d_s}|
      + (k \leftrightarrow j) \\
  & \sim & - \frac{3 \delta_{\lambda'}^{ksp} \delta_{\lambda'}^{jps}}
     {8 \pi^2} \frac{|m_{d_s}m_{d_p}|}{m_{SUSY}}, \\
\end{eqnarray}
 where 
$ \tilde{A}^{{\bf p}{\bf r}}  
=  -[(A \lambda')^{R{\bf p}{\bf r}}v_R + v_u \mu_R 
\lambda^{'R{\bf p}{\bf r}}]$,
and we work in the down quark mass eigenstate 
basis: $\vec{\lambda^{'st}} \cdot \vec{v}$ is diagonal.
 The integrals $I$
are listed in Appendix \ref{Appint}.
 The last line
corresponds to the last column  of table \ref{t2}.
  Note that we do not divide
by 2 when we symmetrise on $i$ and $j$, because
the diagram with $i \leftrightarrow j$ is different.
This agrees with \cite{chunetal,hirschetal}
\footnote{We thank E.J. Chun for a discussion
of this point.} and
disagrees with \cite{GH}. 
 It is clear
in the mass insertion approximation
that one should not divide by 2, because
 in one case $\nu_i$ couples to $\tilde{D}^c$
and in the other case to $\tilde{Q}$. In the
mass eigenstate formalism, one can see
that these are distinct diagrams not included
in the sum over $p,r,s$ by considering the case
where only $\lambda^{'132}$ and $\lambda^{'223}$
are non-zero.

In our previous paper, we were unclear about whether
the charged goldstone boson of SU(2)
could mix with the $\{E^c \}$. This was due
to various sign discrepancies in the literature. As
expected, the goldstone is pure SU(2)---so
there is no loop diagram  propagating a $W^{\mu}$ that is 
proportional to a gauge times a Yukawa coupling.

\section{When are the bilinears important in loops?}
\label{sect3}

Bilinear contributions to loops have traditionally been 
neglected, which intuitively seems reasonable 
if they are ``small'' in a ``sensible''
basis close to the MSSM. However, this apparently reasonable 
oversight is confusing, because the
size of what is neglected depends on the basis choice.
In this section, we provide basis-independent conditions
for when  the bilinears can be neglected in the loops---
assuming the mass matrix
elements are then calculated in a sensible
basis where the bilinears are small.
These are not hard and fast rules, but  estimates
of when the extra bilinear diagrams should be included.

A slightly different approach would be to
consider
the relative size of all the diagrams,
and catalogue all
the permutations of which diagrams should be included
and which neglected. For instance, it is
possible that
the trilinears are negligeable with respect to
the bilinear diagrams,  or that
the $\lambda$ trilinear
contributes only a small correction to the
$\lambda'$ trilinear, and can be ignored.
Here we assume that the trilinear
contributions are always calculated,
 because they take little effort. Then we ask
whether  $\delta_B$ should
also be included---this is a few more
diagrams. Finally
we consider adding the $\delta_{\mu}$ contributions,
which is many more diagrams. We do not
specifically discuss the case where
$\delta_{\mu}$ should be included but
the $\delta_{B}$ contributions are insignificant, because
the $\delta_B$ diagrams are relatively little work
compared  to the many $\delta_{\mu}$ diagrams.

In our previous paper, we outlined three cases:
\begin{itemize}
\item {\bf A:} all the bilinear contributions to
loop diagrams are negligeable.
In this case the loop contributions to
the neutrino mass matrix are the 
usual trilinear   diagrams  of figures  2a  and 2b.
\item  {\bf B:} $\delta_{\mu}$ is negligeably small,
but $\delta_B$ should be included.
In addition to the diagrams of
case A, one should consider  the neutral Grossman-Haber loop
of figure 2c, and  possibly  the $\delta_B$ mass insertion
to figure 2a. 
\item   {\bf C:} 
Include all bilinear $\Rpv$ contributions.
There are additional contributions to diagrams
2a, 2b, and 2c, and a new diagram 2d. 
\end{itemize}

We determine basis-independent criteria for when
the bilinears can be neglected in the loops
by comparing our estimates of the size
of each diagram from the last column of table
\ref{t2}. A diagram  should be
included if its contribution
is of order the trilinear loops---however, we want 
to avoid comparing bilinear
loop corrections of the tree mass
(which are irrelevant), with the trilinear
loops contributing  masses to neutrinos which are massless
at tree level. So we distinguish two
cases : $m^{tree} \lappeq m^{loop}$,
and  $m^{tree} \gg m^{loop}$.

If $m^{tree} \lappeq m^{loop}$, we simply
compare bilinear to trilinear loop mass matrix
elements. Since 
$\delta_{\mu}^i$ is not large enough
to produce  
 $m^{tree} \gappeq  m^{loop}_{ij}$, it
is negligeable in
the loops, so we are in cases A or B.

If  $m^{tree} \gg m^{loop}$,
we consider only the loop mass matrix 
for the neutrinos that are massless
at tree level. 
We compare the bilinear and trilinear
contributions to this sub-matrix.
Cases A, B, or C can arise  if $m_3^{tree} \gg m^{loop}$.

\subsection{case B---including the soft bilinear $B_i$}

We first consider when $\delta_B$
should be included in the loops. The Grossman-Haber loop
is of order\footnote{We are neglecting flavour violation among
the sneutrinos, which could induce a significant loop
mass even if $\vec{B} \simeq b \vec{\mu}$ \cite{sachasteve}.}
$m_{\nu}^{ij} \sim g^2 \delta_B^i \delta_B^jm_{\chi}/(64 \pi^2)$,
which is potentially large because it
is proportional to gauge rather than Yukawa couplings.
Recall that the canonical trilinear diagrams
are of order $m_{\nu}^{ij}\sim  \delta_{\lambda'}^{i33} 
\delta_{\lambda^{'}}^{j33} m_b^2/(8 \pi^2 m_{SUSY} )$.

If   the tree mass is small,  $m^{tree} \lappeq m^{loop}$,
then the Grossman-Haber loop  should be included 
if it is of order the trilinear
loop.
This will occur
if
\beq
\frac{g}{2} \delta_B^j \gappeq \delta_{\lambda'}^{jqp} \sqrt{h_d^q h_d^p} ,
~ 
 \delta_{\lambda}^{jkl}\sqrt{h_e^{k}h_e^{l}},
\eeq
or the condition to neglect $\delta_B$ can
be roughly estimated as
\beq
 \delta_B^j \ll \delta_{\lambda'}^{j33} h_b, 
 \delta_{\lambda}^{j33}h_{\tau} ~~~~~( m^{tree} \lappeq m^{loop}),
\eeq

Alternatively, if  $m^{tree} \gg m^{loop}$,
 we are only interested
in the loop contributions to the
mass of neutrinos $\nu_2$ and $\nu_1$
who are massless at tree level: if $\vec{B}$
is aligned with $\vec{\mu}$, then the GH loop
is a correction to the tree-level mass, and can be neglected
along with other such  loops.
$\vec{B}$ can be decomposed in components parallel
and perpendicular to $\mu$:
\beq
\vec{B} = \vec{B}_{\perp} + \vec{B}_{\parallel},
\label{Bperp}
\eeq
with $\vec{B}_{\parallel} = (\vec{B} \cdot \vec{\mu}) 
\vec{\mu}/| \vec{\mu} |^2$. The part
of the Grossman-Haber loop
proportional to $\delta_{B_{\parallel}}^i \delta_{B_{\parallel}}^j$
is a loop correction to the tree mass $m^{tree}$.
The  $\delta_{B_{\parallel}}^i \delta_{B_{\perp}}^j$
terms mix $\nu^{tree}$ with $\nu_2$ and $\nu_1$, and
the mass matrix from the GH loop for $\nu_1$ and
$\nu_2$ is 
\beq
\mnu \propto
 \frac{ \delta_{B_{\perp}}^i  \delta_{B_{\perp}}^j}{64 \pi^2} m_{\chi}.
\eeq
This will be comparable or greater than the
trilinear loops when
\beq
\frac{g}{2} \delta_{B_{\perp}}^i= 
\frac{g}{2}(\delta_B^i - \frac{\vec{B} \cdot \vec{\mu}}{|\vec{B}|| \vec{\mu}|}
\delta_{\mu}^i)
 \gappeq  \delta_{\lambda'_{\perp}}^{jqp} \sqrt{h_d^q h_d^p} , ~
 \delta_{\lambda}^{jkl}\sqrt{{h_e^{k}h_e^{l}}}.
\eeq
We define $ \delta_{\lambda'_{\perp}}^{jqp} =
\delta_{\lambda'}^{jqp} - \delta_{\mu}^j  
  ( \vec{\lambda'}^{qp} \cdot \vec{\mu})
/| \vec{\mu}|$.
This translates roughly into the condition that
$\delta_B$ can be neglected in the loops if
\beq
\delta_{B_{\perp}}^i \ll 
\delta_{\lambda'_{\perp}}^{i33}h_b, 
 \delta_{\lambda}^{i33}h_{\tau} 
 ~~~~~(m^{tree}_{\nu} \gg m^{loop}_{\nu}).
\eeq

\subsection{case C---including $\mu_i$}

The tree-level neutrino mass is 
$\sim \sum_i (\delta_{\mu}^i)^2 m_{\chi}$,
so if $\delta_{\mu}^i$ is
relevant in the loops, then
$m^{loop} \ll m^{tree}$.
The additional diagrams proportional to $ \delta_{\mu}^i$
should be included if they induce a loop mass
for $\nu_2$ and $\nu_1$  greater than or of order  the trilinear or
GH loops.

We first consider the various  $\delta_{\mu}^i$  contributions
to diagrams $a$ and $d$ of figure \ref{fl}
which are of order
$\sim [\delta_{\mu}^i (h_e^i)^2]
[\delta_{\mu}^j (h_e^j)^2] m_{SUSY}$.
Generically the vector
$(\delta_{\mu}^{\tau}h^{\tau}h^{\tau},$ $\delta_{\mu}^{\mu}h^{\mu}h^{\mu},$
$\delta_{\mu}^e h^eh^e)$
will not be aligned with 
$(\delta_{\mu}^{\tau},\delta_{\mu}^{\mu},\delta_{\mu}^e)$, so
the $\delta_{\mu}$ bilinear contributions
should be included if
\beq
\delta_{\mu}^i (h_e^i)^2 \gappeq \delta_{\lambda}^{jkl} \sqrt{ h_e^k h_e^l},
 \delta_{\lambda'_{\perp}}^{jpq} \sqrt{ h_d^p h_d^q},
~~~~~i,j,k,l,p,q~ {\rm not~summed}.
\label{18}
\eeq
A rough guide to when these $\delta_{\mu}$
corrections can be neglected is therefore
\beq
\delta_{\mu}^j h^{\tau} << \delta_{\lambda}^{i \tau \tau},
   \delta_{\lambda'_{\perp}}^{i 33} \left(\frac{h_b}{h_\tau} \right).
\label{19}
\eeq
There are also  bilinear loop contributions of the form
\beq
[m^{loop}]_{ij} \sim [\delta_{\mu}^i (h_e^i)^2]
[\delta_{\mu}^j ] m_{SUSY}  ~~~(i,j \in \{\tau,\mu,e\}).
\label{above}
\eeq
These initially appear more significant,
because they are only suppressed by two, rather
than four, trilinears or yukawas. However,
$m^{loop}_{\nu} \ll m^{tree}_{\nu}$ and
$\hat{\nu}^{tree} \propto (\delta_{\mu}^{\tau},
\delta_{\mu}^{\mu},\delta_{\mu}^{e})$,
so contributions of the form
of equation (\ref{above}) mix 
$\hat{\nu}^{tree}$ with $\nu^2$ and $\nu^1$.
These latter  two neutrinos, which are massless at tree level,
therefore acquire a seesaw mass of order
\begin{eqnarray}
[m^{loop}]_{ij}& \sim & \sum_k \delta_{\mu}^i (h_e^i)^2  m_{SUSY}
\frac{(\delta_{\mu}^k)^2}{|\delta_{\mu}|^2  m_{SUSY}}
 (h_e^j)^2\delta_{\mu}^j  m_{SUSY} \nonumber \\
& \sim & [\delta_{\mu}^i (h_e^i)^2][  (h_e^j)^2\delta_{\mu}^j]  m_{SUSY}
  ~~~(i,j \in \{3,2,1 \} ).
\label{end}
\end{eqnarray}
This is the mass matrix structure we considered
before equation (\ref{18}), and will make a significant contribution
if equation (\ref{18}) is satisfied. So 
 equation (\ref{18}) is the condition for when $\delta_{\mu}$ 
bilinears should be included in loops, or equivalently,
equation (\ref{19}) is a rough estimate of when the $\delta_{\mu}$
loops can be neglected.

\section{Phenomenology}
\label{pheno}

In this section we briefly consider
the constraints that can be set on $\Rpv$ couplings
from solar and atmospheric neutrino data. We
follow the approach of \cite{AM1,AM2}, who
set bounds on $\Rpv$ models corresponding to case A---models
where the bilinears are negligeable
in loops and neutrino loop masses
are due to trilinear couplings. We discuss here a toy
model where the $\Rpv$ is in
the bilinears, and the $\delta_B$ loops
make a significant contribution to $\mnu$
(case B of the previous section).  We leave
case C---where $\delta_{\mu}$ should be
included in the loops---for a subsequent analysis \cite{AMS}.
We  would like to establish bounds on the different 
set of basis independent combinations
of coupling constants that contribute to the neutrino 
mass matrix. These bounds are more indicative of orders
 of magnitude than steadfast constraints.
In order to proceed we make the following assumptions.
First, we use the results of reference \cite{AM1}, which constrain
a general $3\times 3$ symmetric mass matrix
from neutrino data considering: atmospheric and solar 
neutrino experiments, CHOOZ
experiment and the constraint from neutrinoless double beta decay.
The overall conclusion of this analysis is that the mass matrix elements
should be constrained to be on the order of $| \; \mnu | \lsim 0.1$eV.
Here we present our bounds allowing each individual 
matrix element to take on the maximum value of
0.1 eV. This simple approach can be extended to obtain 
bounds when we include combinations of
contributions to the neutrino mass matrix from 
$\delta_{\mu}^i$, $\delta_{B}^{i}$, $\delta_{\lambda}^{ijk}$, 
for any given model which contains  these R-parity 
violating terms originally in the Lagrangian.

Secondly, we assume that each of the contributions from the diagrams that
we have found is separately the unique contribution at a given time\footnote{This may be incorrect in a poorly chosen basis,
where different diagrams cancel each other \cite{AMS}.}.
In this way we are trying to obtain the {\it weakest}
constraint on the basis-independent combination of 
coupling constants. That is we allow each separate 
contribution to take on its maximum value. We can then choose from all
of the bounds  which is the strongest constraint on
a given  combinations of coupling constants. We 
also assume, unless otherwise indicated,
that all supersymmetric mass scales are the same of order $m_{\mathrm susy}$.
The next approximation we make is that when we sum over neutralinos or
charginos  there is no suppression arising from  
the mixing angles.
We also assume  that the magnitude of all 
R-parity violating couplings constants to be generation
blind separately. That is, all $B_i$ to be of 
the same order of magnitude, and similarly
for $\mu_i$, $\lambda_{ijk}$, $\lambda'_{ijk}$, 
but we do not suppose beforehand that
R-parity violating coupling constants arising 
from different terms in the Lagrangian 
are also of the same order of magnitude.
The numerical results for all diagrams are summarized in appendix \ref{D1}.

As we can see from the  tables in appendix \ref{D1} the strongest bounds
for the combination of basis-independent coupling constants
$\delta_{B}^{i}\delta_{B}^{j}$ arises from the Grossman-Haber diagram:
\beq
\delta_{B}^{i}\delta_{B}^{j} < 3 \times 10^{-10} ~~.
\eeq
For $\delta_{B}^{i}\delta_{\mu}^{j}$
the best bound comes from diagram 19, 
which is the neutral 
Grossman-Haber loop with R-parity violation
at points I and VI:
\beq
\delta_{B}^{i}\delta_{\mu}^{j} < 7 \times 10^{-10} ~~.
\eeq
(We use $\tan \beta = 2$ and a generic
$m_{susy} \sim 100$ GeV  for both
these bounds.)
 The strongest constraint on
 $\delta_{\mu}^{i}\delta_{\mu}^{j}$
is from the tree-level mass:
\beq
\delta_{\mu}^{i}\delta_{\mu}^{j} < 10^{-12} ~~.
\eeq
For comparison, we list the bounds
which can be derived from the remaining diagrams
in Appendix \ref{D1S}.

\subsection{Toy Model}

Suppose we only allow the presence of the  basis-independent
combinations of coupling constants $\delta_{\mu}^{i}$ and  $\delta_{B}^{i}$.
A non-zero value of  $\delta_{\mu}^{i}$ can arise either
from having $\mu_i\neq 0$ or $<\snu_i>\neq 0$,
so this model could arise if all
the $\Rpv$ originates in the soft SUSY breaking terms $B_i$,
which induces a misalignment
between the vev and superpotential couplings. 
(This would induce $\delta_{\lambda'} \sim h_b \delta_{\mu}$
and  $\delta_{\lambda} \sim h_{\tau} \delta_{\mu}$,
which we neglect because we assume our model to be
in case B, where contributions of this size
can be neglected.) This differs slightly from
the usual bilinear model, discussed
in detail in \cite{num,chunetal,hirschetal},
where the $\Rpv$ originates in the GUT-scale
misalignment between $\vec{\mu}$ and the trilinears,
so that $\vec{B}$ becomes misaligned
with respect to $\vec{\mu}$ while running down to
the weak scale.
From the results of appendix \ref{D1}, we  see that the
relevant contributions to the neutrino mass matrix will arise
from the tree level contribution plus diagrams 3 and 19.
The upper bound $\delta_{\mu}^i \lappeq 10^{-6}$
from the tree level mass contribution ensures that
(for low $\tan \beta$) the remaining diagrams,  
 which in principle contribute to the neutrino mass matrix
through mass insertions of $\delta_{\mu}^i$ in
the loops, are negligeable.

With these three contributions we can obtain neutrino masses in
the phenomenologically interesting region which
can simultaneously satisfy atmospheric and solar neutrino data
constraints.
There are several ways to see this. 
The full mass matrix using our simple approximations is given by,
\beq
m_{\nu}^{ij} = m_{\mathrm susy} \delta_{\mu}^{i} \delta_{\mu}^{j} +  a_1  m_{\mathrm susy} \delta_{B}^{i} \delta_{B}^{j} + a_1  m_{\mathrm susy} (\delta_{B}^{i} \delta_{\mu}^{j}
+ \delta_{B}^{j} \delta_{\mu}^{i}),
\label{44}
\eeq
where $a_1 \simeq {g^2\over 64\pi^2 }$.  
Certain subcases of this mass matrix
can generate neutrino mass textures, which have 
been analyzed in the literature,
that can produce neutrino masses and mixing angles 
in accordance with present neutrino data. 
As a simple example, we see that if we set
$\delta_{B}^{3} =0$, and neglect the $\delta_{\mu}
\delta_B$ term in equation (\ref{44}), then the structure of the 
neutrino mass matrix corresponds to
the one  analysed in case 1  of ref. \cite{AM2}. 
In this case the neutrino mass matrix
at one-loop consists simply of the tree-level 
term plus a correction to
the {1-2} submatrix.  The subcase analyzed  in  ref. 
\cite{AM2} had this same structure 
but the loop correction was due to the trilinear diagram.

It is well known that
the tree level contribution can give mass to only one of the
neutrinos.  Together with     the loop contributions one
of the neutrinos will have a mass
\beq
m_{\nu^{3}} =  m_{\mathrm susy} \sum_i 
|\delta^i_{\mu} + \sqrt{a_1} \delta^i_{B_{\parallel}}|^2,
\eeq
where we have used the decomposition 
$\vec{B} = \vec{B}_{\perp} + \vec{B}_{\parallel}$ of
equation (\ref{Bperp}).

The sub-mass-matrix for the remaining
 two neutrinos $\nu^1$ and $\nu^2$
is given by
\beq
m_{\nu}^{ij} = a_1   m_{\mathrm susy}  
\delta_{B_{\perp}}^{i}  \delta_{B_{\perp}}^{j}.
\label{loopij}
\eeq

Terms of the form $ \delta_{B_{\perp}}^{i}  
\delta_{\mu}^{i}$  mix
 the   heavier $\nu^3 $ and the lighter $(\nu^1,\nu^2)$. 
It can be seen from equation (\ref{end}) that
the seesaw mass generated from this mixing is
suppressed, given our choice of SUSY parameters,  
compared to the direct loop contribution given in eq. (\ref{loopij}).
So in this case, this  model corresponds
to case 1 of \cite{AM2}.
This simple example \footnote{ We leave a more detailed 
numerical analysis for future work.} shows that a
 hierarchical neutrino mass spectrum, which 
satisfies atmospheric and solar neutrino data 
constraints, can easily be obtained  by taking 
$ |\delta_{\mu} + \sqrt a_1\delta_{B_{\parallel}}| 
\sim few \times 10^{-6}$ and
 $\delta_{B_{\perp}}^{i} \sim few
\times 10^{-6} - 10^{-5}$. This gives us a massless 
neutrino and two massive neutrinos such
that $\Delta m_{atm} \sim m_{\nu^{3}}^{2}$ and  
$\Delta m_{solar} \sim m_{\nu^{2}}^{2}$. It is 
also possible to obtain other types of spectra for different input values.

\section{Conclusions}
\label{concl}

There are many diagrams which can contribute to the neutrino mass matrix in
the framework of the MSSM without R-parity. In general, in the literature
only a few of these have been considered.
In the present paper we have obtained the full analytic expression for the different diagrams
contributing to the neutrino mass matrix from R-parity violating  bilinear and trilinear coupling constants. We have expressed each contribution
both in the $\snuvi =0$ basis and in terms basis-independent combinations of couplings.
We have also shown when the separate diagrams should be included/are relevant in the context
of a given consistent framework.
We have presented bounds on combinations of the basis-independent couplings constants
from neutrino experimental data, and shown that a simple toy model of bilinear R-parity violation
can successfully accomodate the necessary mass squared differences to account for
the neutrino oscillations.

\section*{Acknowledgements}

We would like to thank Eung Jin Chun,  Herbi Dreiner
and Paolo Gambino for useful conversations. The work of M.L. was partially suported
by Colciencias-BID, under contract no. 120-2000.

\section*{Appendices}

 In these Appendices, we present
 the one loop contributions
 to the neutrino mass matrix,
 {\it excluding} the one-loop
 corrections to the tree level mass, because
 we are interested in the one-loop
 masses for the neutrinos
that  are massless at tree level.
We identify these loops
in Appendix \ref{loop}.
In Appendix \ref{A1}, we list diagrams in the order
of the table 2. We give the amplitude 
for each non-zero diagram in the $\snuvi = 0$
basis.  
In Appendix \ref{B1}, we systematically
list all diagrams, ordering them by the roman
numerals that identify 
in figure \ref{fl} where are the two
units of $\Rpv$. The amplitudes
in Appendix B are ``basis-independent'',
that is they are expressed
in terms of invariants and MSSM parameters.
Notice that the neutrino mass matrix elements
are {\it minus} the amplitude for the diagram,
$\mnu = -{\cal M}_{ij}(p^2 = 0)$, so our
formulae are for $- \mnu$.
We present the $\Rpv$ Feynman rules
in Appendix \ref{C1}. Appendix \ref{D1} contains our numerical results for
bounds placed on combinations of  basis-independent couplings constants
from neutrino data.

\appendix

\section{renormalisation---which are the finite loops}
\label{loop}

We wish to neglect loop corrections
to the tree level mass, 
because we do not need such accuracy and because we prefer
to avoid the issue of renormalisation.  
We therefore neglect all diagrams involving
gauge bosons and those with $\Rpv$ at
I and VIII.  We
initially expected the
remaining loops to be finite, because they could contribute
mass to the neutrinos who are massless at tree level.
 However, diagram 17 (figure 2d with $\Rpv$ at I and VII)
 is  renormalisation scale  ($Q^2$)
dependent.
  In
this Appendix,  we 
show that  the offending $Q^2$ dependent
diagram is a loop
correction to the tree mass.

The tree mass matrix can be written
\beq
\mnu =  m_3^{tree} (\hat{e}_3^{tree})_i (\hat{e}_3^{tree})_j
\eeq
where $\hat{e}_3^{tree}$ is the eigenvector
associated with $ m_3^{tree}$. At one loop,
both the mass and eigenvector are modified:
\beq
\mnu =  (m_3^{tree} + \Delta m_3) (\hat{e}_3^{tree}+ \Delta \hat{e}_3)_i 
(\hat{e}_3^{tree}+ \Delta \hat{e}_3)_j
\eeq
We do not want to include loops
contributing to $ \Delta m_3$ or $ \Delta \hat{e}_3$.
We do not calculate 
gauge loops and diagrams  with $\Rpv$ at
I and VIII, which contribute to
 $ \Delta m_3$.  However,
there are also loop corrections which change the direction
of $\hat{e}_3$, that is
loop corrections to the angles in
the rotation matrix which diagonalises
the $ 7 \times 7$ neutral fermion
mass matrix.  These  loops
are included in our calculation, and 
we  want to identify them  and throw them out.

We work in the mass insertion approximation,
in the flavour basis where $(\hat{L}_{\ell})^J \propto \lambda^{JK \ell}
v_{K}$.  This is the basis where the charged
lepton mass matrix is diagonal in the MSSM.
Some care is required in determining
the  $\{v_I \}$. For one-loop neutrino
masses, it is sufficient to
use the vevs $\{v_I \}$ that
minimise the tree level potential.
To identify the one-loop corrections
to the tree mass, we must use some
one-loop choice for the  $\{v_I \}$.
There are various possibilities, 
for instance we could use the vevs
which minimise the one-loop
effective potential, or we
could use the tree + one-loop
masses that mix neutrinos and $\tilde{h}_d^o$
with gauginos. We opt for the latter,
because in such a basis it is easy
to separate the loop corrections to $m_3 \hat{e}_3^T \hat{e}_3$ from
the loop masses $m_2$ and $m_1$.
So we choose a basis where there
are no tree or one-loop mass terms mixing $\tilde{w}^o$
with $\nu_i$. This means that there
will be small tree-level sneutrino vevs 
which cancel the one-loop
 $\tilde{w}^o  \nu_i$ mass.
These vevs are formally of one loop
order, so are irrelevant
inside the loops, because
their contribution would be of two loop order.
The contribution of these sneutrino vevs 
to the numerical value of
the tree level mass is also a higher
order effect and therefore negligeable.
The one place these vevs must be included
is in the one-loop $\nu_i \tilde{b}_o$ mass,
where the tree level sneutrino vev contribution
(formally one loop order) cancels the scale
dependent part of the one-loop  $\nu_i \tilde{b}_o$ mass,
but leaves a finite  $\nu_i \tilde{b}_o$ mixing.
These can
contribute to the loop masses
$m_2$ and $m_1$, as can loop
contributions to $h_d \nu_i$ mixing.

We know that in the flavour basis at
tree level the neutrino mass is given
by eqn \ref{mtree}.  The one-loop
expression will be the same, in our
present basis, provided we identify
the $\mu_i$ of eqn \ref{mtree} with
the tree+ one-loop mass terms that
mix $\tilde{h}_u^o$ with $\nu_i$.
Since we are only
interested in the lowest order contributions
to neutrino masses, we only
include the tree level contribution
to $m_3$. Knowing the
one-loop corrections allows
us to identify  the finite loops
that generate masses $m_2$ and $m_1$:
these will be loops without mass
insertions on the external legs, and
loops with one mass insertion that
mixes a neutrino with 
$\tilde{h}_d$ or $\tilde{b}^o$.
The loops mixing a neutrino with
$\tilde{h}_d$ or $\tilde{b}^o$
are {\it finite}  in our present basis,
as we show in Appendix \ref{A1}, diagram 17.

In summary, we identify the loops
contributing to one-loop neutrino masses
$m_2$ and $m_1$ to be those without
mass insertions on the external  legs, and
diagrams with a mass insertion
on one of the legs that mixes a neutrino
with the $\tilde{b}^o$ or $\tilde{h}_d^o$
components of a neutralino. The
remaining loop diagrams, with
a mass insertion on both external
legs, or  a mass insertion mixing
$\nu_i$ with $\tilde{w}^o$ or $\tilde{h}_u$
are corrections to the tree mass.

\section{Diagrams in $\snuvi = 0$ basis}
\label{A1}
\label{Appint}

We define 

\beq
\tilde{A}^{{\bf ml}} = - \biggl( (h_e A)^{{\bf ml}} \frac{v_d}{\sqrt 2} + \mu \frac{v_u}{\sqrt 2}h_e^{{\bf ml}}\biggr),
\eeq

and

\beq
\tilde{A}_d^{{\bf ml}} = - \biggl( (h_d A)^{{\bf ml}} \frac{v_d}{\sqrt 2} + 
\mu \frac{v_u}{\sqrt 2}h_d^{{\bf ml}}\biggr).
\eeq

We include $\tilde{A}$ and $\tilde{A}_d$ in the
mass insertion approximation, and seperately
diagonalise the $ 3 \times 3$ slepton ( or down squark) doublet
and singlet mass matrices. Bold face indices
indicate these slepton or squark
mass eigenstate bases, see Appendix C for
details.  Lepton and down
quark mass eigenstate indices are in ordinary type.

We also define

\beq
 I(m_{1}, m_{2 }) = - {1\over 16\pi^2} {m_1^2\over m_1^2 -m_2^2} \ln\frac{m_1^2}{m_2^2},
\eeq

\bea
I(m_{1}, m_{2}, m_{3}) &=&  \int {d^4k\over (2\pi)^4} {1\over k^2 + m_1^2} {1\over k^2 + m_2^2} {1\over k^2 + m_3^2} \nonumber \\
&=& {1\over m_1^2 -m_2^2} \biggl( I(m_{1}, m_{3}) - I
(m_{2}, m_{3})\biggr),
\eea

\bea
I(m_{1}, m_{2}, m_{3}, m_4) &=&{1\over m_1^2 -m_2^2} \biggl[{1\over m_1^2 -m_3^2}\biggl( I(m_{1}, m_{4}) \nonumber \\
& - &I
(m_{3}, m_{4})\biggr) - {1\over m_2^2 -m_3^2}\biggl( I(m_{2}, m_{4}) - I
(m_{3}, m_{4})\biggr)\biggr],
\eea

\bea
\lefteqn{I(m_{1}, m_{2}, m_{3}, m_4, m_5) =} && \nonumber \\
&&{1\over m_1^2 -m_2^2} \biggl[\biggl( I(m_{1}, m_3, m_{4}, m_5) - I
(m_2, m_{3}, m_{4}, m_5)\biggr].
\eea

\begin{itemize}

\item Diagram 1. $R_{p}$-violation at II and VII from $\lambda$ couplings.

\beq
(II, VII)  =  - \sum_{l,{\bf k,m}} \lambda^{il{\bf k}} \lambda^{j{\bf m}l} 
m_{e_{l}} \tilde{A}^{{\bf mk}} I(m_{e_{l}}, m_{\tilde{E}_{{\bf k}}}, 
m_{\tilde{L}_{{\bf m}}})  + (i \leftrightarrow j) \label{lII,VII}
\eeq
which can be rewritten as
\beq
(II, VII)  =  \sum_{l,k} \lambda^{ilk} \lambda^{jkl} m_{e_{l}} m_{e_{k}} (A + \mu_4 \tan\beta) I(m_{e_{l}}, m_{\tilde{E}_{k}}, m_{\tilde{L}_{k}})  + (i \leftrightarrow j)
\eeq
when we take $\tilde{A}^{{\bf mk}}$ to be diagonal.

\item Diagram 2. $R_{p}$-violation at II and VII from $\lambda'$ couplings.

\beq
(II, VII)  =  - 3 \sum_{l,{\bf k,m}} \lambda'^{il{\bf k}} 
\lambda'^{j{\bf m}l} m_{d_{l}} \tilde{A}_{d}^{{\bf mk}} 
I(m_{d_{l}}, m_{\tilde{D}_{{\bf k}}}, m_{\tilde{Q}_{{\bf m}}})  
+ (i \leftrightarrow j)
 \label{lpII,VII}
\eeq
which can be rewritten as
\beq
(II, VII)  = 3 \sum_{l,k} \lambda'^{ilk} \lambda'^{jkl} m_{d_{l}} m_{d_{k}} (A + \mu_4 \tan\beta) I(m_{d_{l}}, m_{\tilde{D}_{k}}, 
m_{\tilde{Q}_{k}})  + (i \leftrightarrow j)
\eeq
when we take $\tilde{A}_d^{mk}$ to be diagonal.

\item Diagram 3. $R_{p}$-violation at (IV,VI) in the Grossman-Haber
diagram. 
\bea (IV,VI)
& = & \sum_{{\alpha},{\bf k,m}} 
\frac{g^{i{\bf k}} B^i g^{j {\bf m}}B^j~ }{4 \cos^2\beta}   
(Z^*_{{\alpha}2} - Z^*_{{\alpha}1} g'/g)^2  m_{\chi_{\alpha}^o}
 \left\{ I(m_h, m_{\tilde{\nu}_{\bf k}},m_{\tilde{\nu}_{\bf m}} 
m_{\chi_{\alpha}^o}) 
\cos^2(\alpha - \beta) \right. \nonumber \\
 && \left. +  
I(m_H, m_{\tilde{\nu}_{{\bf k}}}, m_{\tilde{\nu}_{{\bf m}}}, 
m_{\chi_{\alpha}^o}) \sin^2(\alpha - \beta) 
- I(m_A, m_{\tilde{\nu}_{{\bf k}}},m_{\tilde{\nu}_{{\bf m}}}, 
m_{\chi_{\alpha}^o}) \right\}
 \label{GHIV,VI}
\eea

\item Diagram 4.  $R_{p}$-violation at (I,VII) and (II,VIII) from 
$\lambda'- \lambda'$ 
diagram.

\bea
(I, VII) + (II, VIII) &=& 3 \sum_{\alpha,{\bf l},k,{\bf m}} 
\mu_i {Z^{*}_{\alpha 3} 
Z^{*}_{\alpha 4}\over m_{\chi_{\alpha}}} h_d^{k{\bf l}} m_{d_{k}} 
\lambda'^{j{\bf m}k} \tilde{A}_d^{{\bf ml}}  I(m_{d_{k}}, m_{\tilde{D}_{{\bf l}}}, m_{\tilde{Q}_{{\bf m}}})
\nonumber \\
&&  +3 \sum_{\alpha,{\bf l},k,{\bf m}} 
\mu_j {Z^{*}_{\alpha 3} Z^{*}_{\alpha 4}\over 
m_{\chi_{\alpha}}} h_d^{k{\bf l}} m_{d_{k}} \lambda'^{i{\bf m}k} 
\tilde{A}_d^{{\bf ml}}   \nonumber \\ &&
I(m_{d_{k}}, m_{\tilde{D}_{{\bf l}}}, m_{\tilde{Q}_{{\bf m}}}) 
 \label{lpI,VII}
\eea
which simplifies to

\bea
(I, VII) + (II, VIII) &=& 
-3  \sum_{\alpha,k} \mu_j {Z^{*}_{\alpha 3} Z^{*}_{\alpha 4}\over 
m_{\chi_{\alpha}}} h_d^k m_{d_{k}}^2 \lambda'^{ikk} 
(A + \mu_4 \tan\beta)  \nonumber \\ & & 
 I(m_{d_{k}}, m_{\tilde{D}_{k}}, m_{\tilde{Q}_{k}}) +  (i \leftrightarrow j)
\eea
when both $ h_d^{k{\bf l}}$ and $ \tilde{A}_d^{{\bf ml}}$ are taken to be diagonal.

\item Diagram 5  $R_{p}$-violation at II and VI from $\lambda - 
\lambda$ diagram.

\beq
(II, VI) = \sum_{{\bf l,m}} \lambda^{ij{\bf l}} m_{e_{j}} h_{e}^{jj} 
\tilde{A}^{{\bf ml}} 
B_{\bf m}  \tan\beta 
 I(m_{e_{j}}, m_{\tilde{E}_{{\bf l}}}, m_{\tilde{L}_{{\bf m}}},m_{H^{+}})  
 +  (i \leftrightarrow j)
 \label{lII,VI}
\eeq
which reduces to

\beq
(II, VI) = - \sum_{l} \lambda^{ijl} m_{e_{j}} m_{e_{l}} h_{e}^{j} 
(A +\mu_4\tan\beta) B_l  \tan\beta 
 I(m_{e_{j}}, m_{\tilde{E}_{l}}, m_{\tilde{L}_{l}},m_{H^{+}})  +  (i \leftrightarrow j)
\eeq
when we take 
 $ \tilde{A}^{{\bf ml}}$ diagonal.

\item Diagram 6.  $R_{p}$-violation at (I,VII) and (II,VIII) from  $\lambda - 
\lambda$ diagram.

\bea
(I, VII) + (II, VIII) &=&
  \sum_{\alpha,{\bf l},k,{\bf m}} 
\mu_i {Z^{*}_{\alpha 3} Z^{*}_{\alpha 4} \over 
m_{\chi_{\alpha}^o}} h_e^{k{\bf l}} m_{e_{k}} \lambda^{j{\bf m}k} 
\tilde{A}^{{\bf ml}}  
I(m_{e_{k}}, m_{\tilde{E}_{{\bf l}}}, m_{\tilde{L}_{{\bf m}}})
\nonumber \\
& & + \sum_{\alpha,{\bf l},k,{\bf m}} \mu_j 
{Z^{*}_{\alpha 3} Z^{*}_{\alpha 4}\over 
m_{\chi_{\alpha}^o}} h_e^{k{\bf l}} m_{e_{k}} \lambda^{i{\bf m}k} 
\tilde{A}^{{\bf ml}}  
\nonumber \\ &&I(m_{e_{k}}, m_{\tilde{E}_{{\bf l}}}, 
m_{\tilde{L}_{{\bf m}}}) 
 \label{lI,VII}
\eea
which simplifies to

\bea
(I, VII) + (II, VIII) &=& 
-   \sum_{\alpha,k} \mu_j {Z^{*}_{\alpha 3} Z^{*}_{\alpha 4}\over m_{\chi_{\alpha}}} h_e^k m_{e_{k}}^2 \lambda^{ikk} (A + \mu_4 \tan\beta) \nonumber \\
&& I(m_{e_{k}}, m_{\tilde{E}_{k}}, m_{\tilde{L}_{k}}) +  (i \leftrightarrow j)
\eea
when both $ h_e^{k{\bf l}}$ and $ \tilde{A}^{{\bf ml}}$ 
are taken to be diagonal.

\item Diagram 7. $R_{p}$-violation at I and V from  $\lambda - 
\lambda$ diagram.

\beq
(I, V) =  \sum_{\alpha,{\bf l},k} \mu_i  {Z^{*}_{\alpha 3} 
Z^{*}_{\alpha 4}\over m_{\chi_{\alpha}^o}} h_e^{j{\bf l}} (h_e^{k{\bf l}} 
\mu_k 
{v_u\over \sqrt 2} \sin^2 \beta) h_e^{jj} m_e^j   
I(m_{e_{j}}, m_{\tilde{E}_{{\bf l}}}, m_{H^{+}}) +  (i \leftrightarrow j).
 \label{lI,V}
\eeq
Taking the lepton Yukawa couplings to be diagonal this simplifies to
\beq
(I, V) =  \sum_{\alpha} \mu_i  {Z^{*}_{\alpha 3} 
Z^{*}_{\alpha 4}\over m_{\chi_{\alpha}^o}} (h_e^{j})^{2} 
(h_e^{j} \mu_j {v_u\over \sqrt 2} \sin^2 \beta)  m_e^j   
I(m_{e_{j}}, m_{\tilde{E}_{j}}, m_{H^{+}})+  (i \leftrightarrow j).
\eeq

\item Diagram 8. $R_{p}$-violation at II and V from $\lambda - 
\lambda$ diagram.

\beq
(II, V) = - \sum_{{\bf k}, m} \lambda^{ij{\bf k}} m_{e_{j}} h_{e}^{jj} 
\mu_m h_{e}^{m{\bf k}} {v_u\over \sqrt 2} \sin^2\beta 
I(m_{e_{j}}, m_{\tilde{E}_{{\bf k}}}, m_{H^{+}})+  (i \leftrightarrow j).
 \label{lII,V}
\eeq
which becomes

\beq
(II, V) = - \sum_k \lambda^{ijk} m_{e_{j}} h_{e}^{jj} \mu_k h_{e}^{k} {v_u\over \sqrt 2} \sin^2\beta I(m_{e_{j}}, m_{\tilde{E}_{k}}, m_{H^{+}})+  (i \leftrightarrow j).
\eeq
for diagonal Yukawa couplings.

\item Diagram 9. $R_{p}$-violation at III and V from $\lambda - 
\lambda$ diagram.

\bea
(III,V)& =& \sum_{\alpha,{\bf k},n} {v_u\over \sqrt 2} \sin^2\beta \mu_{n} 
\mu_{j} m_{e_{j}}
h_{e}^{jj} h_{e}^{i{\bf k}} h_{e}^{n{\bf k}} 
V_{\alpha 2}^{*} U_{\alpha 2}^{*} 
m_{\chi_{\alpha}}^{+} \nonumber \\ &&
I(m_{e_{j}}, m_{\tilde{E}_{{\bf k}}}, m_{H^{+}},m_{\chi_{\alpha}^+})+  (i \leftrightarrow j)
 \label{lIII,V}
\eea
which simplifies to

\bea
(III,V)& =& \sum_{\alpha} {v_u\over \sqrt 2} \sin^2\beta \mu_{i} \mu_{j} 
m_{e_{j}} h_{e}^{j} h_{e}^{i^{2}} V_{\alpha 2}^{*} U_{\alpha 2}^{*} 
m_{\chi_{\alpha}}^{+} \nonumber \\ &&
I(m_{e_{j}}, m_{\tilde{E}_{i}}, m_{H^{+}},m_{\chi_{\alpha}^+})+  (i \leftrightarrow j)
\eea
for diagonal Yukawa couplings.

\item Diagram 10. $R_{p}$-violation at III and VIII from $\lambda -
\lambda$ diagram.

\bea
(III,VIII)& =& -\sum_{\alpha,\beta,{\bf k,l},m} h_{e}^{i{\bf k}} 
\tilde{A}^{{\bf lk}} h_{e}^{{\bf l}m} 
m_{e_{m}}\mu_{m} \mu_{j} m_{\chi_{\alpha}^+} 
V_{\alpha 2}^{*} U_{\alpha 2}^{*} \times \nonumber \\& & 
 {Z_{\beta 3}^{*} Z^*_{\beta 4}\over m_{\chi_{\beta}}^{o}} 
I(m_{e_{m}}, m_{\tilde{E}_{{\bf k}}}, m_{\tilde{L}_{{\bf l}}},
m_{\chi_{\alpha}}^{+})+  (i \leftrightarrow j)
 \label{lIII,VIII}
\eea
when we take diagonal Yukawa couplings this reduces to

\bea 
(III,VIII)& =& \sum_{\alpha,\beta,m} (A + \mu_{4}\tan\beta) 
h_{e}^{i^{2}} m_{e}^{i^{2}} \mu_{i} \mu_{j} 
 m_{\chi_{\alpha}^+} 
V_{\alpha 2}^{*} U_{\alpha 2}^{*} \times \nonumber \\& & 
{Z_{\beta 3}^{*} Z^*_{\beta 4}\over m_{\chi_{\beta}}^{o}} I(m_{e_{i}}, m_{\tilde{E}_{i}}, m_{\tilde{L}_{i}}, m_{\chi_{\alpha}}^{+})+  (i \leftrightarrow j)
\eea

\item Diagram 11. $R_{p}$-violation at I and VI from 
$\lambda -
\lambda$ diagram.

\beq
(I,VI) = - \sum_{\alpha,{\bf l,m} } \mu_{i}  {Z_{\alpha 3}^{*} Z^*_{\alpha 4}
\over m_{\chi_{\alpha}^{o}}} h_{e}^{j{\bf l}} h_{e}^{jj} m_{e_{j}} 
\tilde{A}^{{\bf ml}} B_{{\bf m}} \tan\beta  
I(m_{e_{j}}, m_{\tilde{E}_{{\bf l}}}, m_{\tilde{L}_{{\bf m}}},m_{{H}^{-}})+  (i \leftrightarrow j)
 \label{lI,VI}
\eeq

This reduces to

\bea
(I,VI)&  =&  \sum_{\alpha} \mu_{i}  {Z_{\alpha 3}^{*} Z^*_{\alpha 4}
\over m_{\chi_{\alpha}^{o}}} (h_{e}^{jj} m_{e_{j}})^2   
(A + \mu_4 \tan \beta) B_{j} \tan\beta  
\nonumber \\ &&
I(m_{e_{j}}, m_{\tilde{E}_{j}}, m_{\tilde{L}_{j}},m_{{H}^{-}})+  (i \leftrightarrow j)
\eea

with diagonal Yukawa couplings.

\item Diagram 12. $R_{p}$-violation at III and VII  from $\lambda -
\lambda$ diagram.

\beq 
(III,VII) = \sum_{\alpha,{\bf k,l},m} h_{e}^{i{\bf k}} \tilde{A}^{{\bf lk}}
 \lambda^{j{\bf l}m} 
m_{e_{m}} \mu_{m} V_{\alpha 2}^{*} U_{\alpha 2}^{*} m_{\chi_{\alpha}^+} 
I(m_{e_{m}}, m_{\tilde{E}_{{\bf k}}}, m_{\tilde{L}_{{\bf l}}},m_{\chi_{\alpha}^+})+  (i \leftrightarrow j)
 \label{lIII,VII}
\eeq
With diagonal Yukawa couplings this reduces to

\bea
(III,VII)&  =& -\sum_{\alpha,m} h_{e}^{ii} m_{e_i}
(A + \mu_4 \tan \beta) \lambda_{jim} 
m_{e_{m}} \mu_{m} V_{\alpha 2}^{*} U_{\alpha 2}^{*} 
m_{\chi_{\alpha}^{+}} \nonumber \\ &&
I(m_{e_{m}}, m_{\tilde{E}_{i}}, m_{\tilde{L}_{i}},
m_{\chi_{\alpha}^{+}})+  (i \leftrightarrow j).
\eea

\item Diagram 13. $R_{p}$-violation at III and VI  from $\lambda -
\lambda$ diagram.

\bea
(III,VI) &=& - \sum_{\alpha,{\bf k,l}} 
h_{e}^{i{\bf k}} \tilde{A}^{{\bf lk}} B_{{\bf l}}\tan\beta 
h_{e}^{jj} m_{e_{j}} \mu_{j} \nonumber \\
& & V_{\alpha 2}^{*} U_{\alpha 2}^{*} m_{\chi_{\alpha}}^{+}  
I(m_{e_{j}}, m_{\tilde{E}_{{\bf k}}}, 
m_{\tilde{L}_{{\bf l}}}, m_{H^+},m_{\chi_{\alpha}^{+}})+  (i \leftrightarrow j)
 \label{lIII,VI}
\eea
which reduces to

\bea 
(III,VI) &=& \sum_{\alpha} (A + \mu_{4}\tan\beta) h_{e}^{i} m_{e_{i}}  
h_{e}^{j} m_{e_{j}}
B_{i} \mu_{j}  \nonumber \\
&&V_{\alpha 2}^{*} U_{\alpha 2}^{*} m_{\chi_{\alpha}^{+}}
  I(m_{e_{j}}, m_{\tilde{E}_{i}}, m_{\tilde{L}_{i}}, 
m_{H^+},m_{\chi_{\alpha}^{+}})+  (i \leftrightarrow j).
\eea

\item Diagram 14. $R_{p}$-violation at III and IV in $g - \lambda$ loop.

\beq
(III,IV) = -\sum_{\alpha, {\bf l}} {g^{i {\bf l}}
\over \sqrt 2} U^*_{\alpha 1} V^*_{\alpha 2}
 m_{\chi_{\alpha}}^{+} \mu_{j} m_{e_{j}} h_{e}^{jj} B_{{\bf l}} \tan\beta 
I(m_{e_{j}}, m_{\tilde{L}_{{\bf l}}}, m_{H^+},m_{\chi_{\alpha}^{+}})
+  (i \leftrightarrow j)
 \label{glIII,IV}
\eeq
which reduces for $g^{i {\bf l}} = g$ to

\beq
(III,IV) = -\sum_{\alpha} {g\over \sqrt 2} U^*_{\alpha 1} V^*_{\alpha 2}
 m_{\chi_{\alpha}}^{+} \mu_{j} m_{e_{j}} h_{e}^{jj} B_{i} \tan\beta 
I(m_{e_{j}}, m_{\tilde{L}_{i}}, m_{H^+},m_{\chi_{\alpha}^{+}})
+  (i \leftrightarrow j)
\eeq

\item Diagram 15. $R_{p}$-violation at III and VIII in $g - \lambda$ loop.

\beq
(III,VIII) = -\sum _{\alpha \beta,k,  {\bf l}} 
{g^{i  {\bf l}}\over \sqrt 2} h_{e}^{{\bf l}k} 
m_{e_{k}} \mu_{k} \mu_{j} V^*_{\alpha 2} U^*_{\alpha 1} 
m_{\chi_{\alpha}}^{+} {Z_{\beta 3}^{*} Z^{*}_{\beta 4}\over 
m_{\chi_{\beta}}^{o}} I(m_{e_{k}}, m_{\tilde{L}_{{\bf l}}}, 
m_{\chi_{\alpha}^+})+  (i \leftrightarrow j)
 \label{glIII,VIII}
\eeq

When the charged slepton mass eigenstates are aligned with
the charged lepton mass eigenstates, this becomes

\beq
(III,VIII) = -\sum _{\alpha \beta} {g\over \sqrt 2} h_{e}^{i} 
m_{e_{i}} \mu_{i} \mu_{j} V^*_{\alpha 2} U^*_{\alpha 1} 
m_{\chi_{\alpha}}^{+} {Z_{\beta 3}^{*} Z^{*}_{\beta 4}\over 
m_{\chi_{\beta}}^{o}} I(m_{e_{i}}, m_{\tilde{L}_{i}}, 
m_{\chi_{\alpha}^+})+  (i \leftrightarrow j)
\eeq

\item Diagram 16 $R_{p}$-violation at I and III in $g - \lambda$ loop.

There are two possible diagrams with $\Rpv$ at I and III.
Firstly the neutralino can arrive as a $\tilde{h}_d$
at vertex II, where a $\tilde{w}^-$ is
absorbed. Secondly
the neutralino can arrive as a gaugino, in which case
the $\tilde{h}_d^-$ part
of the chargino is absorbed.

\bea
(I,III)& =& - \sum _{\alpha, \beta} {g\over \sqrt 2} \sin^{2}
\beta h_{e}^{jj} m_{e_{j}} \mu_{i} \mu_{j} V_{\alpha 2}^{*} U_{\alpha 1}^{*} 
m_{\chi_{\alpha}^+} {Z_{\beta 3}^{*} Z^{*}_{\beta 4}\over m_{\chi_{\beta}^o}} 
I(m_{e_{j}}, m_{H^+}, m_{\chi_{\alpha}^+})
\nonumber \\
&&  + \sum _{\alpha, \beta} {g\over \sqrt 2} \sin^{2}
\beta h_{e}^{jj} m_{e_{j}} \mu_{i} \mu_{j} V^{*}_{\alpha 2} U^{*}_{\alpha 2} 
m_{\chi_{\alpha}^+} {(Z_{\beta 2}^{*} + Z_{\beta 1}^{*}g'/g)
 Z^{*}_{\beta 3}\over m_{\chi_{\beta}^o}} \times
\nonumber \\ &&
I(m_{e_{j}}, m_{H^+}, m_{\chi_{\alpha}^+})
+  (i \leftrightarrow j)
 \label{glI,III}
\eea

\item Diagram 17 $R_{p}$-violation at I and VII in $g - \lambda$ loop.
We also include here the   $g - \lambda'$ loop of similar form,
 with sleptons/leptons replaced
by squarks and quarks. 
This diagram,
with the $\Rpv$ at I removed, corresponds to
a mass mixing $ \tilde{w}^o  \nu $, 
or $ \tilde{b} \nu $. 

We  want to show
that in the basis where  $ \tilde{w}^o  \nu $ 
masses are zero at one-loop, the
 $ \tilde{b}  \nu $ mass is finite.
To do this, we compute both contributions
in the  $\snuvi = 0$ basis. The basis
we want is where the tree level
 $ \tilde{w}^o  \nu $ mass ($= g \snuvi/2$)
cancels the loop  $ \tilde{w}^o  \nu $  mass---
so in this basis the  $ \tilde{b}  \nu $ mass
is $- g' \snuvi/2 + $ loop = $ \frac{g'Z_{\alpha 1}^*}{gZ_{\alpha 2}^*}($ the
 $ \tilde{w}^o  \nu $ loop) +
the  $ \tilde{b}^o  \nu $ loop, which
 is finite.


$\tilde{w}^o  \nu $ mixing gives:

\bea
(I,VII)& =& - \sum _{\alpha, k,{\bf l}} 
{g^{k{\bf l}} \over \sqrt 2} m_{e_{k}} 
\mu_{i}
 \frac{Z_{\alpha 3}^* Z^{*}_{\alpha 2}  }
{ m_{\chi_{\alpha}^{o}}} \lambda^{{\bf l}jk} 
 B_0(0, m_{e_{k}}, m_{\tilde{L}_{\bf l}})~~~~~~~~~~~~~~~~  
  \nonumber \\ &&
 - 3 \sum _{\alpha, k,{\bf l}} {g^{k{\bf l}} \over \sqrt 2} m_{d_{k}} 
\mu_{i}
 \frac{Z_{\alpha 3}^* Z^{*}_{\alpha 2}  }
{ m_{\chi_{\alpha}^{o}}} \lambda^{'j{\bf l}k} 
 B_0(0, m_{d_{k}}, m_{\tilde{Q}_{\bf l}})+  (i \leftrightarrow j)
 \label{glI,VII}
\eea

and $\tilde{b} - \nu$ mixing gives

\bea
(I,VII)& =& - \sum _{\alpha k,{\bf l}} m_{e_{k}} 
\mu_{i}
 \frac{Z_{\alpha 3}^* Z^{*}_{\alpha 1}  }
{ m_{\chi_{\alpha}^{o}}}  
\left\{  {g^{'k{\bf l}}\over \sqrt 2} \lambda^{{\bf l}jk} 
 B_0(0, m_{e_{k}}, m_{\tilde{L}_{\bf l}}) 
-2  {g^{'k{\bf l}}\over \sqrt 2} \lambda^{kj{\bf l}} 
 B_0(0, m_{e_{k}}, m_{\tilde{E}_{\bf l}}) \right\}
  \nonumber \\ &&
- \sum _{ \alpha k,{ \bf l}} m_{d_{k}} 
\mu_{i}
 \frac{Z_{\alpha 3}^* Z^{*}_{\alpha 1}  }
{ m_{\chi_{\alpha}^{o}}}  
\left\{  {g^{'k{\bf l}}\over \sqrt 2} \lambda^{'j{\bf l}k} 
 B_0(0, m_{d_{k}}, m_{\tilde{Q}_{\bf l}}) 
-2  {g^{'k{\bf l}}\over \sqrt 2} \lambda^{'jk{\bf l}} 
 B_0(0, m_{d_{k}}, m_{\tilde{D}_{\bf l}}) \right\}
  \nonumber \\ && ~~~ +  (i \leftrightarrow j)
 \label{17bnu}
\eea

Using  $B_0(p^2 = 0, m_1, m_2) = I(m_2,m_1)  
- \ln (m_1^2/Q^2) +1$, one can see that
the finite contribution to  $\tilde{b} - \nu$ mixing,
which can contribute to the loop neutrino masses $m_2$ and
$m_1$ is the sum of the above two expressions
(\ref{glI,VII},\ref{17bnu})
with  $B_0(p^2 = 0, m_1, m_2) \rightarrow  I(m_2,m_1) $.

(Notice that  the $g^{'k{\bf l}}$ appearing
in different terms have absorbed different
rotation matrices, so are not the same. See equation
(\ref{BglI,VII})  for formulae with explicit diagonalisation matrices.)

We can set $g^{k{\bf l}} = g$, $g^{'k{\bf l}} = g'$
and $k = {\bf l}$ in the above expression if 
the sfermion and fermion  mass eigenstate bases
are the same.

\item  Diagram 18 $R_{p}$-violation at III and VII in $g - \lambda$ loop;
as noted in ref. \cite{DL1}, 
this is zero if the sleptons are mass-degenerate.

\beq
(III,VII) = \sum _{\alpha, k,{\bf l}} {g^{i {\bf l}}
\over \sqrt 2} m_{e_{k}} \mu_{k}
V^{*}_{\alpha 2} U^{*}_{\alpha 1}
 m_{\chi_{\alpha}^{+}} \lambda^{{\bf l}jk} 
 I(m_{e_{k}}, m_{\tilde{L}_{\bf l}}, 
m_{\chi_{\alpha}^{+}} )+  (i \leftrightarrow j)
 \label{glIII,VII}
\eeq

which reduces for  $g^{k{\bf l}} = g$ to
\beq
(III,VII) = \sum _{\alpha, k} {g\over \sqrt 2} m_{e_{k}} \mu_{k}
V^{*}_{\alpha 2} U^{*}_{\alpha 1}
 m_{\chi_{\alpha}^{+}} \lambda^{ijk} 
 I(m_{e_{k}}, m_{\tilde{L}_i}, m_{\chi_{\alpha}^{+}} )+  (i \leftrightarrow j)
\eeq

\item  Diagram 19 $R_{p}$-violation at I and VI in Grossman-Haber loop.

We neglected this diagram in our previous paper---the expression
for the amplitude is lengthly. In the interaction eigenstate
basis, the incident neutrino
can turn into an up-type higgsino/down-type higgsino 
(gaugino),  which subsequently turns into a
 gaugino and up/down Higgs
(an up- or down-type Higgs and higgsino) at the vertex II.
In mass eigenstate basis, these possibilities generate
a number of terms:

\begin{eqnarray}
(I,VI) + (IV,VIII)  &= &\sum _{\alpha, \beta,{\bf l}}
 {g g^{j {\bf l}} \over  4 \cos \beta} 
 \mu_{i} {
Z^*_{\alpha 3} \over m_{\chi_{\alpha}^{o}}}  m_{\chi_{\beta}^{o}}
( Z^*_{\beta 2} -  Z^*_{\beta 1} g'/g) B_{\bf l}  \times \nonumber \\ 
&&\left\{- \left[ 
 Z^*_{\alpha 4}( Z_{\beta 2}^*-  Z_{\beta 1}^* g'/g) \sin \alpha +
   ( Z^*_{\alpha 2}-  Z_{\alpha 1}^* g'/g)Z_{\beta 3}^* \cos \alpha
 \nonumber \right. \right.\\  &&\left.
~~~
+ ( Z^*_{\alpha 2}-  Z^*_{\alpha 1} g'/g)  Z_{\beta 4}^*\sin \alpha \right]
  \cos(\alpha - \beta) 
   I(m_h,m_{\chi_{\beta}^{o}}, m_{\tilde{\nu}_{\bf l}})\nonumber  \\ &&
+ \left[  Z^*_{\alpha 4}( Z_{\beta 2}^*-  Z_{\beta 1}^* g'/g) \cos \alpha
 -( Z^*_{\alpha 2}-  Z_{\alpha 1}^* g'/g)  Z_{\beta 3}^*\sin \alpha
 \nonumber \right.\\ && \left.
~~~ +( Z^*_{\alpha 2}-  Z^*_{\alpha 1} g'/g) Z_{\beta 4}^* \cos \alpha \right]
   \sin(\alpha - \beta) 
  I(m_H,m_{\chi_{\beta}^{o}}, m_{\tilde{\nu}_{\bf l}})\nonumber \\ && 
- \left[   Z^*_{\alpha 4}( Z_{\beta 2}^*-  Z_{\beta 1}^* g'/g) \sin \beta
 + ( Z^*_{\alpha 2}-  Z_{\alpha 1}^* g'/g)Z_{\beta 3}^* \cos \beta
  \nonumber \right.\\ && \left.  \left.
~~~
+ ( Z^*_{\alpha 2}-  Z^*_{\alpha 1} g'/g)  Z_{\beta 4}^*\sin \beta \right] 
I(m_A,m_{\chi_{\beta}^{o}}, m_{\tilde{\nu}_{\bf l}}) 
\right\}  + (i \leftrightarrow j)
 \label{GHI,VI}
\eea

\item Diagram 20 $\Rpv$ at I and V with $g - \lambda$ couplings. This
differs from figure 2d): the internal line is a higgsino and
there is an $\tilde{A}$ mass insertion on the scalar line.

\bea
(I,V)& =& 
- \sum_{\alpha,\beta,{\bf m}}
\mu_i {g \over \sqrt{2}} { Z^*_{\alpha 3} 
\over m_{\chi_{\alpha}^o}} \left\{ ( Z^*_{\alpha 2} + 
Z^*_{\alpha 1} g'/g)  
U^*_{\beta 2} V^*_{\beta 2}  + 
 Z^*_{\alpha 3} 
U^*_{\beta 2} V^*_{\beta 1} \right\}   m_{\chi_{\beta}^+} \nonumber \\& &
m_{e_{\bf m}} h_e^{j{\bf m}}  \mu_{\bf m} \sin^2 \beta \tan \beta 
I(m_{\chi_{\beta}^+}, m_{\tilde{E}_{\bf m}}, m_{H^+}) + (i \leftrightarrow j)
 \label{glI,V}
\eea
which reduces for diagonal yukawas to 

\bea
(I,V)& =& 
- \sum_{\alpha,\beta}
\mu_i {g \over \sqrt{2}} { Z^*_{\alpha 3} 
\over m_{\chi_{\alpha}^o}} \left\{  ( Z^*_{\alpha 2} + 
Z^*_{\alpha 1} g'/g)
U^*_{\beta 2} V^*_{\beta 2}  +
Z^*_{\alpha 3} 
U^*_{\beta 2} V^*_{\beta 1}
\right\} m_{\chi_{\beta}^+} \nonumber \\& &
m_{e_j} h_e^{j}  \mu_j \sin^2 \beta \tan \beta \times
I(m_{\chi_{\beta}^+}, m_{\tilde{E}_j}, m_{H^+}) + (i \leftrightarrow j)
\eea

\item Diagram 21 $\Rpv$ at I and IV with $g - \lambda$ couplings.

The particle identities do not correspond
to the labels on figure 2 d). The $\nu_i$ can mix with
a higgsino, then the loop fermion is a chargino,
arriving at VII as  $\tilde{h}_d$. Alternatively,
the  $\nu_i$ turns into
a gaugino, then there are
two possibilities. Firstly, the loop fermion can
be a higgsino, arriving at II as
$\tilde{h}_u$ and
at VII as $\tilde{h}_d$. Secondly
the loop fermion can be
a charged lepton, and the scalar arriving at VII is a charged Higgs.

\bea
(I,IV) & =& 
\sum_{\beta,\alpha,{\bf p,m}}
\mu_i {g \over \sqrt{2}} { Z^*_{\alpha 3}
\over m_{\chi_{\alpha}^o}}  
\left\{  ( Z^*_{\alpha 2} + 
Z^*_{\alpha 1} g'/g)
U^*_{\beta 2} V^*_{\beta 2}  
\right\}
  m_{\chi_{\beta}^+} \nonumber \\ &&   
 h_e^{j{\bf m}} \tilde{A}^{{\bf pm}} B_{\bf p}
\tan \beta   
I(m_{\chi_{\beta}^+}, m_{\tilde{E}_{\bf m}}, m_{\tilde{L}_{\bf p}}, m_{H^+}) 
+ (i \leftrightarrow j)
\nonumber \\ &&
+ \sum_{\alpha, {\bf l}}
\mu_i {g^{j {\bf l}} \over \sqrt{2}} { Z^*_{\alpha 3} ( Z^*_{\alpha 2} +
Z^*_{\alpha 1} g'/g)
\over m_{\chi_{\alpha}^o}}  m_{e_j} h_e^{jj} \tan \beta  B_{\bf l}\times
\nonumber \\ &&
I(m_{e_j}, m_{\tilde{L}_{\bf l}}, m_{H^+})  + (i \leftrightarrow j)
 \label{glI,IV}
\eea

which for diagonal yukawas reduces to 

\bea
(I,IV) & =& - 
\sum_{\alpha, \beta}
\mu_i {g \over \sqrt{2}} { Z^*_{\alpha 3} 
\over m_{\chi_{\alpha}^o}} 
\left\{  ( Z^*_{\alpha 2} + 
Z^*_{\alpha 1} g'/g)
U^*_{\beta 2} V^*_{\beta 2}  
\right\}    m_{\chi_{\beta}^+}  \times \nonumber \\ &&  
 h_e^{j} m_{e_j}(A + \mu_4 \tan \beta)  B_j
\tan \beta 
I(m_{\chi_{\beta}^+}, m_{\tilde{E}_j}, m_{\tilde{L}_j}, m_{H^+}) 
+ (i \leftrightarrow j) \nonumber \\ &&  
+ \sum_{\alpha}
\mu_i {g \over \sqrt{2}} { Z^*_{\alpha 3} ( Z^*_{\alpha 2} 
+ Z^*_{\alpha 1} g'/g)
\over m_{\chi_{\alpha}^o}}  m_{e_j} h_e^{jj} \tan \beta  B_j \times
\nonumber \\ &&
I(m_{e_j}, m_{\tilde{L}_j}, m_{H^+})  + (i \leftrightarrow j)
\eea

\end{itemize}


\section{Basis-independent diagram amplitudes}
\label{B1}

In this appendix we write the contributions
to the neutrino mass matrix $\mnu$ from each diagram
in terms of MSSM parameters and invariants.
The latter can be evaluated in any basis.

We list all imaginable diagrams, and explain
why some are zero or negligeable. 
We write the MSSM parameters in  the  mass eigenstate basis,
with diagonal quark and charged lepton mass matrices. 
See Appendix C for definitions of the matrices
$Z,U,V,L,E,Q$ and $D$ which respectively
diagonalise neutralinos, negative charginos, positive charginos, 
doublet charged sleptons, singlet sleptons, down-type doublet quarks
and singlet down quarks.

We define 

\beq
\tilde{A}^{ml} = - \hat{L}^m_M 
\biggl( (\lambda A)^{IMl} \frac{v_I}{\sqrt 2} + 
\mu_I \frac{v_u}{\sqrt 2} \lambda^{IMl}\biggr),
\eeq
with $\hat{L}^m = \lambda^m \cdot \vec{v}/| \lambda^m \cdot \vec{v}|$,
see equation (\ref{basis}). Similarly we define
\beq
\tilde{A}^{ml}_d = -  
\biggl( (\lambda' A)^{Iml} \frac{v_I}{\sqrt 2} + 
\mu_I \frac{v_u}{\sqrt 2} \lambda^{'Iml}\biggr).
\eeq
Notice that the indices on
$\tilde{A}$ and $\tilde{A}_d$
are in the charged lepton and down quark 
mass eigenstate basis respectively.

\subsection{ $\lambda - \lambda$ diagrams---figure 2a)}

\begin{itemize}

\item  $R_{p}$-violation at (I,VIII): this
is a loop correction to the tree mass, so can be neglected.

\item  $R_{p}$-violation at (I,VII) and (II,VIII)
(diagram 6, equation \ref{lI,VII}).
\bea
(I, VII) + (II, VIII) &=&  \sum_{\alpha,l,k,m,p,n,q} \delta_{\mu}^i
\delta_{\lambda}^{qjn} m_{e^n} h_e^{n} | \mu|  \tilde{A}^{kp} 
 E_{nl} E^*_{pl} L_{qm} L^*_{km} \nonumber \\
& &
 {Z^{*}_{\alpha 3} Z^*_{\alpha 4}\over m_{\chi_{\alpha}^o}} 
  I(m_{e_{n}}, m_{\tilde{E}_{l}}, m_{\tilde{L}_{m}})  +  (i \leftrightarrow j),
\eea
which simplifies to

\bea
(I, VII) + (II, VIII) &=& 
-   \sum_{\alpha,k} \delta_{\mu}^j  \delta_{\lambda}^{kik}
|\mu|  h_e^k m_{e_{k}}^2 (A + |\mu| \tan\beta) \nonumber \\ &&  
{Z^{*}_{\alpha 3} Z^{*}_{\alpha 4}\over m_{\chi_{\alpha}^o}}
I(m_{e_{k}}, m_{\tilde{E}_{k}}, m_{\tilde{L}_{k}}) +  (i \leftrightarrow j),
\eea
when  $ \tilde{A}, L$ and $E$  are taken to be diagonal.

\item $R_{p}$-violation at (I, VI) (diagram 11,  equation \ref{lI,VI})

\bea
(I,VI)& =& - \sum_{\alpha,l,p,r,q} \delta_{\mu}^{i}   \delta_B^{m}
(h_{e}^{j})^2 m_{e_{j}} \tilde{A}^{qp} |B| |\mu| 
E_{jl} E^*_{pl}L_{mr} L^*_{qr} \tan\beta  \nonumber \\ &&
{Z_{\alpha 3}^{*} Z^{*}_{\alpha 4}\over m_{\chi_{\alpha}^{o}}} 
I(m_{e_{j}}, m_{\tilde{E}_{l}}, m_{\tilde{L}_{r}},m_{H^-})
+  (i \leftrightarrow j),
\eea
 for diagonal $\tilde{A},L,E$  this simplifies to 
\bea
(I,VI) &= &\sum_{\alpha} \delta_{\mu}^{i}   \delta_B^{j}
(h_{e}^{j}  m_{e_{j}})^2 (A + |\mu| \tan \beta) |B| |\mu|  \tan\beta 
{Z_{\alpha 3}^{*} Z^{*}_{\alpha 4}\over m_{\chi_{\beta}^{o}}} 
\nonumber \\ &&
I(m_{e_{j}}, m_{\tilde{E}_{j}}, m_{\tilde{L}_{j}},m_{H^-})
+  (i \leftrightarrow j).
\eea

\item $R_{p}$-violation at (I,V) (diagram 7,  equation \ref{lI,V})

\bea
(I, V)& = &- \sum_{\alpha,l,k} \delta_{\mu}^i  |\mu|^2
 (h_e^{j})^2 m_e^j m_e^{k} \delta_{\mu}^k  \sin^2 \beta \tan \beta
E_{jl} E^*_{kl}
{Z^{*}_{\alpha 3} Z^{*}_{\alpha 4}\over m_{\chi_{\alpha}}} 
I(m_{e_{j}}, m_{\tilde{E}_{l}}, m_{H^{+}}) 
\nonumber \\
&&+  (i \leftrightarrow j).
\eea
Taking $\tilde{A},L$ and $E$ to be diagonal this simplifies to

\beq
(I, V) = - \sum_{\alpha} \delta_{\mu}^i  \delta_{\mu}^j |\mu|^2
 (m_e^j h_e^{j})^{2}
 {Z^{*}_{\alpha 3} Z^{*}_{\alpha 4}\over m_{\chi_{\alpha}}} 
 \sin^2 \beta \tan \beta    
I(m_{e_{j}}, m_{\tilde{E}_{j}}, m_{H^{+}})+  (i \leftrightarrow j).
\eeq

\item  $R_{p}$-violation at (I,IV). This is 
not possible, because there is no   $R_{p}$-violating
mass involving  two $E^c$.

\item  $R_{p}$-violation at (I,II).  This is not possible;
$\Rpv$ at II requires three incident leptons.

\item $R_{p}$-violation at (II, VIII). This is included
with  $R_{p}$-violation at (I, VII).

\item $R_{p}$-violation at (II, VII) from $\lambda$ couplings (diagram 1
 equation \ref{lII,VII})

\beq
(II, VII)  =  - \sum_{l,m,n,r,p,q,s} 
\delta_{\lambda}^{inm} \delta_{\lambda}^{rjn} 
m_{e_{n}} \tilde{A}^{sq} E_{mp} E^*_{pq}  L^*_{sl} L_{rl} 
I(m_{e_{n}}, m_{\tilde{E}_{p}}, m_{\tilde{L}_{l}})  + (i \leftrightarrow j),
\eeq
which can be rewritten as
\beq
(II, VII)  =  \sum_{l,k} \delta_{\lambda}^{ilk} \delta_{\lambda}^{jkl}
 m_{e_{l}} m_{e_{k}} (A + |\mu| \tan\beta) 
I(m_{e_{l}}, m_{\tilde{E}_{k}}, m_{\tilde{L}_{k}})  + (i \leftrightarrow j),
\eeq
when we take $\tilde{A}, L$ and $E$ to be diagonal.

\item  $R_{p}$-violation at (II, VI) (diagram 5,  equation \ref{lII,VI})

\bea
(II, VI)&  = & \sum_{m,n,s,p,q,r} \delta_{\lambda}^{ijn}  \delta_{B}^r |B|
m_{e_{j}} h_{e}^{j}  \tilde{A}^{pq}  \tan\beta
  E_{ns} E^*_{qs}  L_{pm}^* L_{rm} \nonumber \\ && 
 I(m_{e_{j}}, m_{\tilde{E}_{s}}, 
m_{\tilde{L}_{m}},m_{H^{+}})  +  (i \leftrightarrow j),
\eea
which reduces to

\bea
(II, VI) & = &- \sum_{l} \delta_{\lambda}^{ijl}  \delta_{B}^l |B|
 m_{e_{j}} m_{e_{l}} h_{e}^{j}  (A + |\mu|\tan\beta)  \tan\beta 
\nonumber \\ && 
I(m_{e_{j}}, m_{\tilde{E}_{l}}, m_{\tilde{L}_{l}},m_{H^{+}})  
+  (i \leftrightarrow j),
\eea
when we take 
 $ \tilde{A}, L $ and $E$  diagonal.

\item $R_{p}$-violation at (II, V) (diagram 8,  equation \ref{lII,V})

\beq
(II, V) = - \sum_{k,l,m} \delta_{\lambda}^{ijk}  \delta_{\mu}^m |\mu|
m_{e_{j}} h_{e}^{j}  m_{{e}_{m}} \tan \beta \sin^2\beta 
E_{kl} E^*_{ml}
I(m_{e_{j}}, m_{\tilde{E}_{l}}, m_{H^{+}})+  (i \leftrightarrow j),
\eeq
which becomes

\beq
(II, V) = - \sum_k \delta_{\lambda}^{ijk} \delta_{\mu}^k |\mu|
m_{e_{j}} h_{e}^{j} m_{e^k}  \tan \beta \sin^2\beta 
I(m_{e_{j}}, m_{\tilde{E}_{k}}, m_{H^{+}})+  (i \leftrightarrow j),
\eeq
for diagonal matrices.

\item $R_{p}$-violation at (II, IV) is 
not possible because there is no $\Rpv$ mass between
two $E^c$.

\item $R_{p}$-violation at (II, III) is not possible; 
$\Rpv$ at II requires three incident leptons.

\item $R_{p}$-violation at (III, VIII) (diagram 10,  equation \ref{lIII,VIII})

\bea
(III,VIII) &=& - \sum_{\alpha,\beta,k,l,m,q,p} 
\delta_{\mu}^{m} \delta_{\mu}^{j} 
 | \mu|^2
h_{e}^{i} \tilde{A}^{pq} h_{e}^{m} m_{e_{m}}
E_{ik} E^*_{qk} L^*_{pl} L_{ml}
 V_{\alpha 2}^{*} U_{\alpha 2}^{*} m_{\chi_{\alpha}^+}  \nonumber \\
&&
{Z_{\beta 3}^{*} Z^{*}_{\beta 4}\over m_{\chi_{\beta}^o}} 
I(m_{e_{m}}, m_{\tilde{E}_{k}}, m_{\tilde{L}_{l}},m_{\chi_{\alpha}^+})+  
(i \leftrightarrow j),
\eea
when we take diagonal $\tilde{A}, L$ and $E$ this reduces to

\bea 
(III,VIII)& =& \sum_{\alpha,\beta} 
 \delta_{\mu}^{i} \delta_{\mu}^{j} | \mu|^2
(A + |\mu| \tan\beta) (h_{e}^{i} m_{e}^{i})^2 
V_{\alpha 2}^{*} U_{\alpha 2}^{*}  m_{\chi_{\alpha}^+}  \nonumber \\
&& {Z_{\beta 3}^{*} Z^{*}_{\beta 4}\over m_{\chi_{\beta}^o}} 
I(m_{e_{i}}, m_{\tilde{E}_{i}}, m_{\tilde{L}_{i}}, 
m_{\chi_{\alpha}^+})+  (i \leftrightarrow j).
\eea

\item $R_{p}$-violation at (III,VII) (diagram 12, equation \ref{lIII,VII})

\bea
(III,VII)& =& \sum_{\alpha,k,l,m,n,q,p}  \delta_{\lambda}^{jnm}
\delta_{\mu}^m |\mu|
h_{e}^{i} \tilde{A}^{qp} m_{e_{m}}  V_{\alpha 2}^{*} U_{\alpha 2}^{*} 
m_{\chi_{\alpha}^+} 
L_{nl} L^*_{ql} E_{ik} E^*_{pk} \nonumber \\ &
&I(m_{e_{m}}, m_{\tilde{E}_{k}}, m_{\tilde{L}_{l}},
m_{\chi_{\alpha}^+})+  (i \leftrightarrow j),
\eea

which becomes, when $\tilde{A}, L $ and $E$ are diagonal

\bea
(III,VII)& =& -\sum_{\alpha,m}  \delta_{\lambda}^{jim}
\delta_{\mu}^m  |\mu|
h_{e}^{i}  m_{e_{m}} m_{e_{i}} (A + |\mu| \tan \beta)
 V_{\alpha 2}^{*} U_{\alpha 2}^{*} 
m_{\chi_{\alpha}^+}  \nonumber \\
&& I(m_{e_{m}}, m_{\tilde{E}_{i}}, m_{\tilde{L}_{i}},
m_{\chi_{\alpha}^+})+  (i \leftrightarrow j).
\eea

\item $R_{p}$-violation at (III, VI) (diagram 13,  equation \ref{lIII,VI}).

\bea
(III,VI)& =& - \sum_{\alpha,k,l,m,q,p} \delta_{\mu}^j \delta_{B}^m |\mu| |B|
h_{e}^{i} \tilde{A}^{qp} \tan\beta h_{e}^{j} m_{e_{j}}
 V_{\alpha 2}^{*} U_{\alpha 2}^{*} m_{\chi_{\alpha}}^{+}  
E_{ik} E^*_{pk} L^*_{ql} L_{ml} 
\nonumber \\& &
I(m_{e_{j}}, m_{\tilde{E}_{k}}, m_{\tilde{L}_{l}}, m_{H^+},
m_{\chi_{\alpha}^+})+  (i \leftrightarrow j),
\eea
which reduces to

\bea 
(III,VI)& =& \sum_{\alpha} \delta_{\mu}^j \delta_{B}^i |\mu| |B|
 (A + |\mu|\tan\beta) h_{e}^{i} m_{e_{i}}  h_{e}^{j} m_{e_{j}}
 V_{\alpha 2}^{*} U_{\alpha 2}^{*} m_{\chi_{\alpha}}^{+} 
\nonumber \\ && I(m_{e_{j}}, m_{\tilde{E}_{i}}, m_{\tilde{L}_{i}}, 
m_{H^+},m_{\chi_{\alpha}^+})+  (i \leftrightarrow j),
\eea

for diagonal $\tilde{A},L$ and $E$.

\item $R_{p}$-violation at (III,  V) (diagram 9,  equation \ref{lIII,V}). 

\bea
(III,V)& =& \sum_{\alpha,k,n}  \delta_{\mu}^{n} \delta_{\mu}^{j} |\mu|^2
\tan \beta m_e^n  \sin^2\beta m_{e_{j}}
h_{e}^{j} h_{e}^{i} 
E_{ik} E^*_{nk}
 V_{\alpha 2}^{*} U_{\alpha 2}^{*} m_{\chi_{\alpha}^+} \nonumber
\\ & & 
I(m_{e_{j}}, m_{\tilde{E}_{k}}, m_{H^{+}},m_{\chi_{\alpha}^+})
+  (i \leftrightarrow j),
\eea
which simplifies to

\bea
(III,V)& =& \sum_{\alpha} \delta_{\mu}^{i} \delta_{\mu}^{j} |\mu|^2
 \tan \beta \sin^2\beta  m_{e_i}m_{e_{j}}
h_{e}^{j} h_{e}^{i} V_{\alpha 2}^{*} U_{\alpha 2}^{*} m_{\chi_{\alpha}}^{+} \nonumber \\ &&
I(m_{e_{j}}, m_{\tilde{E}_{i}}, m_{H^{+}},m_{\chi_{\alpha}^+})
+  (i \leftrightarrow j),
\eea
for diagonal $E$.

\item $R_{p}$-violation at (III, IV), is not
possible because there is no $\Rpv$ mass involving
two $E_c$.

\item $R_{p}$-violation at (IV , VIII) is 
not possible, for the same reason as (III, IV)---
No diagram with $E^c$ leaving VII and arriving
at V can have 
 $R_{p}$-violation at IV, so no
diagrams with $\Rpv$ at (IV,...) are possible.

\item $R_{p}$-violation at (V, VIII) is the same as  (I, V).

\item $R_{p}$-violation at (V, VII) is the same as  (II, V).

\item $R_{p}$-violation at (V, VI) is not possible, because
we cannot put two units of lepton number violation on
one charged line.

\item $R_{p}$-violation at (VI, VIII) is 
not possible because $H^+$ has no $R_p$ conserving
mass with $E^c$.

\item $R_{p}$-violation at (VI, VII) is 
not possible because $H^+$ has no $R_p$ conserving
mass with $E^c$.

\end{itemize}

\subsection{$\lambda'-\lambda'$ diagrams---figure 2b)}

\begin{itemize}

\item $R_{p}$-violation at (I,VIII) is a loop
correction to the tree mass.

\item $R_{p}$-violation at (I,VII) and (II,VIII) from $\lambda'$ couplings
(diagram 4,  equation \ref{lpI,VII}).

\bea
(I, VII) + (II, VIII) &=&
3 \sum_{\alpha,l,k,m,q,p,r} \delta_{\mu}^i |\mu|
{Z^{*}_{\alpha 3} Z^{*}_{\alpha 4}\over m_{\chi_{\alpha}}} h_d^{k} m_{d_{k}} 
\delta_{\lambda'}^{jrk} 
D_{kl} D^*_{ql} Q^*_{pm} Q_{rm} \tilde{A}_d^{pq} 
  \nonumber \\ && 
I(m_{d_{k}}, m_{\tilde{D}_{l}}, m_{\tilde{Q}_{m}})
\nonumber \\ && 
   +3 \sum_{\alpha,l,k,m,q,p,r} \delta_{\mu}^j  |\mu| 
{Z^{*}_{\alpha 3} Z^{*}_{\alpha 4}
\over m_{\chi_{\alpha}}} h_d^{k} m_{d_{k}} \delta_{\lambda'}^{irk} 
D_{kl} D^*_{ql} Q^*_{pm} Q_{rm} \tilde{A}_d^{pq} 
\nonumber \\
&& I(m_{d_{k}}, m_{\tilde{D}_{l}}, 
m_{\tilde{Q}_{m}}),
\eea
which simplifies to

\bea
(I, VII) + (II, VIII) &=&
-3  \sum_{\alpha,k} \delta_{\mu}^j  |\mu| {Z^{*}_{\alpha 3} 
Z^{*}_{\alpha 4}\over m_{\chi_{\alpha}}} h_d^k m_{d_{k}}^2 
\delta_{\lambda'}^{ikk} (A + |\mu| \tan\beta) \times 
 \nonumber \\ && I(m_{d_{k}}, 
m_{\tilde{D}_{k}}, m_{\tilde{Q}_{k}}) +  (i \leftrightarrow j),
\eea
when  $  \tilde{A}_d, Q$ and $D$ are taken to be diagonal.

\item $R_{p}$-violation at II and VII from $\lambda'$ couplings (diagram 2,
 equation \ref{lpII,VII}).

\bea
(II, VII) & = &  - 3 \sum_{l,k,m,n,p,r,s} \delta_{\lambda'}^{ilp} 
\delta_{\lambda'}^{jnl} m_{d_{l}} \tilde{A}_{d}^{sr} 
D_{pk} D^*_{rk} L_{nm} L^*_{sm} \times \nonumber \\ &&
I(m_{d_{l}}, m_{\tilde{D}_{k}}, m_{\tilde{Q}_{m}})  + (i \leftrightarrow j),
\eea
which can be rewritten as
\beq
(II, VII)  = 3 \sum_{l,k} \delta_{\lambda'}^{ilk} \delta_{\lambda'}^{jkl} 
m_{d_{l}} m_{d_{k}} (A + |\mu| \tan\beta) I(m_{d_{l}}, m_{\tilde{D}_{k}}, m_{\tilde{Q}_{k}})  + (i \leftrightarrow j),
\eeq
when we take $\tilde{A}_d, Q$ and $D$ to be diagonal.

\item  Lepton number violation is not possible inside
a squark loop, so the only additional effect that
bilinear $\Rpv$ can have is to induce squark flavour
violation at V. We neglect this because the squark flavour violation
is small: $\delta_{\mu}^i \delta_{\lambda'}^{ipq} |\mu| v_u$.
If we take $m_{\nu} \lappeq $ eV then this 
implies $\delta_{\mu}^i \lappeq 10^{-5}$,
so  $\delta_{\mu}^i \delta_{\lambda'}^{ipq} |\mu| v_u
\lappeq \delta_{\lambda'}^{ipq}$ GeV$^2$, which is negligeable.
We differ here  from reference \cite{KONG2}, where
this flavour violation due to the bilinears is included,
but bilinear lepton number violation is not.

\end{itemize}

\subsection{the Grossman-Haber diagrams---figure 2c)}

\begin{itemize}

\item  $R_{p}$-violation at (I,VIII) is a loop correction to
the tree level mass, so negligeable.

\item  $R_{p}$-violation at (I,VI) and (IV,VIII)
 (diagram 19, equation \ref{GHI,VI}). The basis independent version
of this diagram can be read off from equation (\ref{GHI,VI}),
substituting $\delta_B^n |B|$ for $B^n$, and
 $\delta_{\mu}^i |\mu|$ for $\mu^i$.

\item  $R_{p}$-violation at (IV,VI)  (diagram 3, equation \ref{GHIV,VI}) 
\bea (IV,VI)
& = & \sum_{{\alpha},l,m,k,n} 
\frac{g^2 \delta_{B}^m \delta_{B}^n~  |B|^2}{4 \cos^2\beta} 
  (Z^*_{{\alpha}2} - Z^*_{{\alpha}1} g'/g)^2  m_{\chi_{\alpha}^o}
L^*_{il}L_{ml} L^*_{jk}L_{nk}  \nonumber \\
 &&
\left\{ I(m_h, m_{\tilde{\nu}_l}, m_{\tilde{\nu}_k}, m_{\chi_{\alpha}^o}) 
\cos^2(\alpha - \beta) +  
I(m_{H},  m_{\tilde{\nu}_l}, m_{\tilde{\nu}_k},
 m_{\chi_{\alpha}^o}) \sin^2(\alpha - \beta)  \right. \nonumber \\
 && \left.
- I(m_A, m_{\tilde{\nu}_l}, m_{\tilde{\nu}_k}, m_{\chi_{\alpha}^o}) \right\},
\eea

which for diagonal $L$ becomes
\bea (IV,VI)
& = & \sum_{{\alpha}} 
\frac{g^2 \delta_{B}^i \delta_{B}^j |B|^2 }{4 \cos^2\beta} 
  (Z^*_{{\alpha}2} - Z^*_{{\alpha}1} g'/g)^2  m_{\chi_{\alpha}^o} 
\left\{ I(m_h, m_{\tilde{\nu}_i}, m_{\tilde{\nu}_j}, m_{\chi_{\alpha}^o}) 
\cos^2(\alpha - \beta) \right. \nonumber \\
 && \left. +  
I(m_{H},  m_{\tilde{\nu}_i}, m_{\tilde{\nu}_j},
 m_{\chi_{\alpha}^o}) \sin^2(\alpha - \beta) 
- I(m_A, m_{\tilde{\nu}_i}, m_{\tilde{\nu}_j}, m_{\chi_{\alpha}^o}) \right\}.
\eea

\end{itemize}

\subsection{$g - \lambda$ diagrams---figure 2d)}
\begin{itemize}

\item $R_{p}$-violation at (I, VIII) would be a loop
correction to the tree level mass, so is negligeable.

\item $R_{p}$-violation at (I,VII) (diagram 17, equation \ref{glI,VII}).
See comments before equation \ref{glI,VII}.


\bea
(I,VII)& =& 
 - \sum _{\alpha, k, l,m} {g^{'}\over \sqrt 2} m_{e_{k}} 
\delta_{\mu}^{i} |\mu|
 \frac{Z_{\alpha 3}^* Z^{*}_{\alpha 1}  }
{ m_{\chi_{\alpha}^{o}}}  
\left\{ 2\delta_{\lambda}^{mjk}  L^*_{kl} L_{ml}  
 I( m_{\tilde{L}_{ l}}, m_{e_{k}}) \right.  \nonumber \\ &&
 ~~~ \left.-2 \delta_{\lambda}^{kjm}  E^*_{kl} E_{ml} 
 I( m_{\tilde{E}_{\bf l}}, m_{e_{k}}) \right\}
  \nonumber \\ &&
- \sum _{\alpha, k, l,m} {g^{'}\over \sqrt 2} m_{d_{k}} 
\delta_{\mu}^{i} |\mu|
 \frac{Z_{\alpha 3}^* Z^{*}_{\alpha 1}  }
{ m_{\chi_{\alpha}^{o}}}  
\left\{ 2 \delta_{\lambda^{'}}^{jmk}  Q^*_{kl} Q_{ml} 
 	I( m_{\tilde{Q}_{ l}}, m_{d_{k}})   \right. \nonumber \\ &&
  ~~~ \left.-2 \delta_{\lambda'}^{jkm}  D^*_{kl} D_{ml} 
 I( m_{\tilde{D}_{ l}}, m_{d_{k}}) \right\}
 +  (i \leftrightarrow j)
 \label{BglI,VII}
\eea

The simplifications when $E,L,Q$ and $D$ are
diagonal are obvious, so we do not repeat
these expressions.

\item $\Rpv$ at (I,V) (diagram 20, equation \ref{glI,V}). The particle
labels in figure 2d) do not
apply in this case. The incident $\nu_i$ mixes with a
gaugino, and the internal fermion line is a higgsino,
or $\nu_i$ mixes with $\tilde{h}_u$ and meets
$\tilde{w}^+$ at II. 
The scalar incident at VII is an $E^c$.

\bea
(I,V) & = &
- \sum_{\alpha,\beta,l,m}
\delta_{\mu}^i \delta_{ \mu}^m |\mu|^2
\frac{g}{\sqrt{2}} { Z^*_{\alpha 3}
\over m_{\chi_{\alpha}^o}} \left\{  ( Z^*_{\alpha 2} + Z^*_{\alpha 1} g'/g)
U^*_{\beta 2} V^*_{\beta 2}  + Z^*_{\alpha 3}
U^*_{\beta 2} V^*_{\beta 1}   \right\} \times \nonumber \\ &&
  m_{\chi_{\beta}^+}
m_{e_m} h_e^{j}  \sin^2 \beta \tan \beta
E_{jl} E^*_{ml} 
I(m_{\chi_{\beta}^+}, m_{\tilde{E}_l}, m_{H^+}) + (i \leftrightarrow j).
\eea

This reduces, when $E$ is diagonal to

\bea
(I,V) & =& 
- \sum_{\alpha,\beta}
\delta_{\mu}^i \delta_{ \mu}^j |\mu|^2
\frac{g}{\sqrt{2}} { Z^*_{\alpha 3}
\over m_{\chi_{\alpha}^o}}  
\left\{  ( Z^*_{\alpha 2} + Z^*_{\alpha 1} g'/g)
U^*_{\beta 2} V^*_{\beta 2}  + Z^*_{\alpha 3}
U^*_{\beta 2} V^*_{\beta 1 }\right\}
   m_{\chi_{\beta}^+}  \nonumber \\ &&
m_{e_j} h_e^{j} \sin^2 \beta \tan \beta
I(m_{\chi_{\beta}^+}, m_{\tilde{E}_j}, m_{H^+}) + (i \leftrightarrow j).
\eea

\item $\Rpv$ at (I,IV) (diagram 21,  equation \ref{glI,IV}).
 The neutrino can mix with a gaugino, then there are
two possibilities, which we list separately. Firstly,
the internal
fermion line can be  a higgsino. An $\tilde{A}$ insertion
on the scalar line ensures that the incident
scalar at VII is an $E^c$.  The second possibility is for the
gaugino to  interact
with $\ell, \tilde{L}$ at II. In this case $H^+$ and
$e^c$ are incident at VII.  A third possibility 
is for the incident neutrino
$\nu_i$  could also mix with $\tilde{h}_u$, which meets
a $\tilde{w}$ at II,
and emits a $H_u^+$. (An $\tilde{A}$ mass insertion
on the scalar line, and $m_{\chi^+}$ on the
fermion line, ensure that $E^c$ and $\tilde{h}_d$
arrive at VII. ) This third possibility
is a loop correction to the tree
mass (the loop, without
the external mass insertion, is a contribution
to $\mu_i \nu_i \tilde{h}_u^o$), so we do not include it.

\bea
(I,IV)_1&  = & 
\sum_{\beta,\alpha,l,m,p,q,r} \delta_B^r \delta_{\mu}^i |\mu| |B|
 \frac{g}{\sqrt{2}} { Z^*_{\alpha 3} ( Z^*_{\alpha 2} + Z^*_{\alpha 1} g'/g)
\over m_{\chi_{\alpha}^o}}  
 h_e^{j} E_{jl} E_{ml}^*
\tilde{A}^{pm} L_{rq} L_{pq}^* 
\tan \beta  \times   \nonumber \\ &&
U^*_{\beta 2} V^*_{\beta 2}   m_{\chi_{\beta}^+}
I(m_{\chi_{\beta}^+}, m_{\tilde{E}_l}, m_{\tilde{L}_q}, m_{H^+}) 
+ (i \leftrightarrow j).
\eea

The second possibility gives

\bea
(I,IV)_2 & = &
 \sum_{\alpha,m,n}
\delta_{\mu}^i  \delta_B^n |\mu| |B|
 \frac{g}{\sqrt{2}} { Z^*_{\alpha 3} ( Z^*_{\alpha 2} + Z^*_{\alpha 1} g'/g)
\over m_{\chi_{\alpha}^o}}  m_{e_j} h_e^{j} \tan \beta
L_{jm} L^*_{nm}  \nonumber \\ &&
I(m_{e_j}, m_{\tilde{L}_m}, m_{H^+}) + (i \leftrightarrow j),
\eea

which reduces, for diagonal $\tilde{A},L$ and $E$ to

\bea
(I,IV) & = &
- \sum_{\beta,\alpha} \delta_B^j \delta_{\mu}^i |\mu| |B|
 \frac{g}{\sqrt{2}} { Z^*_{\alpha 3}
\over m_{\chi_{\alpha}^o}} \left\{ ( Z^*_{\alpha 2} + Z^*_{\alpha 1} g'/g) 
U^*_{\beta 2} V^*_{\beta 2}  \right\}  \times  \nonumber \\ &&
  m_{\chi_{\beta}^+}
 h_e^{j} m_{e_j}
(A + |\mu| \tan \beta)
\tan \beta 
 I(m_{\chi_{\beta}^+}, m_{\tilde{E}_j}, m_{\tilde{L}_j}, m_{H^+}) 
\nonumber \\ &&
+ \sum_{\alpha}
\delta_{\mu}^i  \delta_B^j |\mu| |B|
 \frac{g}{\sqrt{2}} { Z^*_{\alpha 3} ( Z^*_{\alpha 2} + Z^*_{\alpha 1} g'/g)
\over m_{\chi_{\alpha}^o}}  m_{e_j} h_e^{j} \tan \beta
 \nonumber \\ &&
I(m_{e_j}, m_{\tilde{L}_m}, m_{H^+}) + (i \leftrightarrow j).
\eea

\item $R_{p}$-violation at I and III (diagram 16, equation \ref{glI,III})

\bea
(I,III)& = &-\sum _{\alpha \beta}  \delta_{\mu}^{i} \delta_{\mu}^{j}
 |\mu|^2 {g\over \sqrt 2} \sin^{2}\beta 
h_{e}^{j} m_{e_{j}}
\left\{ V^*_{\alpha 2} U^*_{\alpha 1} m_{\chi_{\alpha}^+} 
 {Z_{\beta 3}^{*} 
Z^*_{\beta 4}
\over m_{\chi_{\beta}^o}} 
\right. \nonumber \\ && \left.
-V^{*}_{\alpha 2} U^{*}_{\alpha 2} m_{\chi_{\alpha}^+} 
{(Z_{\beta 2}^{*} + Z_{\beta 1}^{*}g'/g)
 Z^{*}_{\beta 3}\over m_{\chi_{\beta}^o}}
\right\}  
 I(m_{e_{j}}, 
m_{H}^{+}, m_{\chi_{\alpha}}^{+})+  (i \leftrightarrow j).
\eea

\item $R_{p}$-violation at III and VIII (diagram 15, equation 
\ref{glIII,VIII})

\bea
(III,VIII)& =& - \sum_{\alpha \beta,k,l} 
\delta_{\mu}^{k} \delta_{\mu}^{j} |\mu|^2{g\over \sqrt 2} 
L^*_{il} L_{kl}
h_{e}^{k} m_{e_{k}}
V^{*}_{\alpha 2} U^{*}_{\alpha 1} m_{\chi_{\alpha}}^{+} 
{Z_{\beta 3}^{*} Z^{*}_{\beta 4}\over m_{\chi_{\beta}}^{o}} 
\nonumber \\ &&
I(m_{e_{k}}, m_{\tilde{L}_{l}}, m_{\chi_{\alpha}^+})+  (i \leftrightarrow j),
\eea
which reduces for diagonal $L$ to

\bea
(III,VIII)& = &
 -\sum_{\alpha \beta} 
\delta_{\mu}^{i} \delta_{\mu}^{j} |\mu|^2 {g\over \sqrt 2} 
h_{e}^{i} m_{e_{i}} 
V^{*}_{\alpha 2} U^{*}_{\alpha 1} m_{\chi_{\alpha}^+}
  \nonumber \\ &&
{Z_{\beta 3}^{*} Z^{*}_{\beta 4}\over m_{\chi_{\beta}}^{o}} 
I(m_{e_{i}}, m_{\tilde{L}_{i}}, m_{\chi_{\alpha}^+})+  (i \leftrightarrow j).
\eea

\item   $R_{p}$-violation at III and VII in $g - \lambda$ loop;
 this is
zero of the sleptons are mass-degenerate (diagram 18, equation \ref{glIII,VII}).

\beq
(III,VII) = \sum_{\alpha, k,l,m} \delta_{ \mu}^{k} |\mu|
 \delta_{\lambda}^{mjk} 
{g\over \sqrt 2} m_{e_{k}} 
V^{*}_{\alpha 2} U^{*}_{\alpha 1}
 m_{\chi_{\alpha}^{+}} L^{*}_{il} L_{ml}
 I(m_{e_{k}}, m_{\tilde{L}_l}, m_{\chi_{\alpha}^{+}} )+  (i \leftrightarrow j),
\eeq

which for diagonal $L$ reduces to

\beq
(III,VII) = \sum_{\alpha, k} \delta_{ \mu}^{k}   |\mu|
\delta_{\lambda}^{ijk} 
{g\over \sqrt 2} m_{e_{k}} 
V^{*}_{\alpha 2} U^{*}_{\alpha 1}
 m_{\chi_{\alpha}^{+}} 
 I(m_{e_{k}}, m_{\tilde{L}_i}, m_{\chi_{\alpha}^{+}} )+  (i \leftrightarrow j).
\eeq

\item $R_{p}$-violation at (III, IV) (diagram 14, equation \ref{glIII,IV})

\bea
(III,IV)& =&  -\sum_{\alpha,l,k} \delta_{\mu}^{j} \delta_{B}^{k}
|\mu| |B|
{g\over \sqrt 2} U^{*}_{\alpha 1} V^{*}_{\alpha 2} 
m_{\chi_{\alpha}^+} m_{e_{j}} h_{e}^{j} 
L^*_{il} L_{kl}\tan\beta \nonumber \\ && I(m_{e_{j}}, 
m_{\tilde{L}_{l}}, m_{H^+},m_{\chi_{\alpha}^+})+  (i \leftrightarrow j),
\eea
which reduces for diagonal $L$ to

\bea
(III,IV)& =& -\sum_{\alpha} \delta_{\mu}^{j} \delta_{B}^{i}|\mu| |B|
{g\over \sqrt 2} U^{*}_{\alpha 1} V^{*}_{\alpha 2} 
m_{\chi_{\alpha}^+} m_{e_{j}} h_{e}^{j} 
\tan\beta \nonumber \\& & I(m_{e_{j}}, 
m_{\tilde{L}_{i}}, m_{H^+},m_{\chi_{\alpha}^+})+  (i \leftrightarrow j).
\eea

\item (IV, VIII)  and (IV, VII) do not exist, because a
doublet slepton must come out of II, so no
particles carrying lepton number would arrive at VII.

\item (IV, V) and  (IV, VI) do not exist, because
there cannot be two units of lepton number violation
on a charged line.

\item (V, VIII)  and (V, VII) do not exist, because a
doublet slepton must come out of II, so no
particles carrying lepton number would arrive at VII.

\end{itemize}

\section{Conventions and Feynman rules}
\label{C1}

We take all coupling constants to be real.
Indices on Yukawa couplings and $A$ are in order doublet($d$)
singlet ($s$): $h_e^{ds}$. We work in the mass eigenstate
basis of the charged leptons and down-type quarks, where
the Yukawa couplings are diagonal and  only need one index.
In Appendix A, we absorb the slepton
and squark mixing
matrices into the Yukawa and gauge couplings
at the vertices, so
they are not diagonal when a scalar meets a fermion.

The MSSM neutralino mass matrix is diagonalised
by the matrix $Z_{m f}$, with flavour eigenstate
index $f$:1..4 corresponding to
$(-i\tilde{B},-i \tilde{W}^o, \tilde{h}_u^o, \tilde{h}_d^o)$.
 The first index $m$ is the mass eigenstate
index. 
The chargino mass matrix is diagonalised by
matrices $U$ and $V$: $ U^* M V^{\dagger} = M_{diag}$,
so the positive [negative] mass eigenstates are $\chi^+_m =
V_{mf} \psi_f^+$ [$\chi^-_m =
U_{mf} \psi_f^-$] , where $ \psi_f^+ = (-i \tilde{w}^+, \tilde{h}_u^+)$ 
and  $ \psi_f^- = (-i \tilde{w}^-, \tilde{h}_d^-)$.
The doublet and singlet charged slepton (down-type quark)
mass matrices are separately
diagonalised by matrices $L_{fm}$ and $E_{fm}$ ($Q_{fm}$ and $D_{fm}$), 
with index order flavour eigenstate--mass eigenstate.  We include
the $A$ term mixing between doublets and singlets
in perturbation theory. The matrices $L,E,Q$ and $D$
do not appear in Appendix A; instead we absorb
them into $\tilde{A},\tilde{A}_d$ and the gauge
and yukawa vertex couplings. We
use bold face indices to indicate the
squark/slepton mass eigenstates in Appendix A, so
\beq
\tilde{A}^{\bf{lm}} = L^*_{p { l}} \tilde{A}^{pq}
 E^*_{q { m}}, ~~~ \tilde{A}^{\bf{lm}}_d = Q^*_{p { l}} \tilde{A}_d^{pq}
 D^*_{q {m}}, ~~~ h_e^{{\bf l} m} =  L_{k{ l}} h_e^{k m} 
\eeq
We also absorb the diagonalisation matrices
into the gauge couplings, so $g$ also appears with
indices. Usually 
\beq
g^{i {\bf l}} = g  L^{ i {l}}
\eeq
however,  $g$ appears in the formulae for diagram 17
having absorbed $L, E,Q$ or $D$.

We use MSSM  Feynman rules from \cite{H+K}, with
additional Feynman rules to include
the $\Rpv$ interactions  in figures
3,4 and 5. Line direction is superfield
chirality.

For a Lagrangian
\beq
{\cal L} = \bar{\psi}( i\partial\! \! \!  / - m) \psi,
\eeq
we fix the phase of the mass insertion
to be $-i$ because
\beq
\frac{i}{{p\! \!  /~ - m}} = \frac{i}{p\! \!  /~} +
\frac{i}{p\! \!   /~} (-i m)\frac{i}{p\! \!  /~}
+ ...
\eeq
The $-i \mu_i$ mass insertions on the external legs of the diagram
mix the incident flavour eigenstate neutrino $\nu_i$
with the $\tilde{h}_u^o$ component
of a neutralino. So
the external leg propagator delivering
a flavour eigenstate neutralino $f$ to
vertex II is:
\beq
\frac{i}{p\!  \!  /~} (-i \mu_i) \sum_{\alpha}
\frac{iZ^*_{\alpha 3} Z^*_{\alpha f}}
{p\! \!  /~ - m_{\chi_{\alpha}^o}}
= \frac{i}{p\! \!  /~}
(-\mu_i) \sum_{\alpha} \left(\frac{Z^*_{\alpha 3} Z^*_{\alpha f}}
{ m_{\chi_{\alpha}^o}} \right),
\eeq
where we have used $p^2 = 0$.  This gives the Feynman
rules of figure 3.

\begin{figure}[htb]
\unitlength1mm
\SetScale{2.8}
\begin{center}
\begin{picture}(150,30)(0,0)
\ArrowLine(0,0)(15,0)
\ArrowLine(30,0)(15,0)
\ArrowLine(30,0)(45,0)
\Text(-2,0)[r]{$\nu_i$}
\Text(47,0)[l]{$\tilde{h}_d$}
\Text(15,5)[c]{$\mu_i$}
\Text(15,0)[c]{x}
\Text(22,-7)[c]{$\tilde{h}_u$}
\Text(30,5)[c]{$m_{\chi_{\alpha}^o}$}
\Text(30,0)[c]{x}
\Text(58,0)[c]{$=$}
\Text(68,0)[r]{$\nu_i$}
\ArrowLine(70,0)(85,0)
\ArrowLine(85,0)(100,0)
\Text(102,0)[l]{$\tilde{h}_d$}
\Text(85,5)[c]{$\mu_i$}
\Text(85,0)[c]{x}
\Text(120,0)[l]{$- \mu_i {Z^*_{\alpha 4}Z^*_{\alpha 3} 
  \over m_{\chi_{\alpha}^o}}$} 
\end{picture}
\begin{picture}(150,30)(0,0)
\ArrowLine(0,0)(15,0)
\ArrowLine(30,0)(15,0)
\ArrowLine(30,0)(45,0)
\Text(-2,0)[r]{$\nu_i$}
\Text(47,0)[l]{$\tilde{w}^o$}
\Text(15,5)[c]{$\mu_i$}
\Text(15,0)[c]{x}
\Text(22,-7)[c]{$\tilde{h}_u$}
\Text(30,5)[c]{$m_{\chi_{\alpha}^o}$}
\Text(30,0)[c]{x}
\Text(58,0)[c]{$=$}
\Text(68,0)[r]{$\nu_i$}
\ArrowLine(70,0)(85,0)
\ArrowLine(85,0)(100,0)
\Text(102,0)[l]{$\tilde{w}^o$}
\Text(85,5)[c]{$\mu_i$}
\Text(85,0)[c]{x}
\Text(120,0)[l]{$- \mu_i {Z^*_{\alpha 3}Z^*_{\alpha 2}
  \over m_{\chi_{\alpha}^o}}$} 
\end{picture}
\begin{picture}(150,30)(0,0)
\ArrowLine(0,0)(15,0)
\ArrowLine(30,0)(15,0)
\ArrowLine(30,0)(45,0)
\Text(-2,0)[r]{$\nu_i$}
\Text(47,0)[l]{$\tilde{b}^o$}
\Text(15,5)[c]{$\mu_i$}
\Text(15,0)[c]{x}
\Text(22,-7)[c]{$\tilde{h}_u$}
\Text(30,5)[c]{$m_{\chi_{\alpha}^o}$}
\Text(30,0)[c]{x}
\Text(58,0)[c]{$=$}
\Text(68,0)[r]{$\nu_i$}
\ArrowLine(70,0)(85,0)
\ArrowLine(85,0)(100,0)
\Text(102,0)[l]{$\tilde{b}^o$}
\Text(85,5)[c]{$\mu_i$}
\Text(85,0)[c]{x}
\Text(120,0)[l]{$- \mu_i {Z^*_{\alpha 3}
Z^*_{\alpha 1}  \over m_{\chi_{\alpha}^o}}$} 
\end{picture}
\end{center}
\vspace{1cm}
\label{extlegs}
\caption{Feynman rules for mass insertions on external
legs. The  left hand column is a more correct
representation; we use the abbrieviated notation 
of the central column in figure \ref{fl}.}
\end{figure}

The $R_p$ violating $ \mu_i$ and $B_i$ mass insertions,
and the $R_p$ conserving $\tilde{A}$ mass
insertion,
 on internal charged lines  are negative
(by SU(2) antisymmetric
contraction), so
effectively appear in the amplitude
with a negative sign:
\beq
\frac{i}{p  \! \!  \!  /~ - m_1} (i|\mu_i|)
\frac{i}
{p  \! \! \!  /~ - m_2} = 
\frac{i}{p  \! \!  \!  /~ - m_1} 
\frac{-|\mu_i|}
{p  \! \! \!  /~ - m_2}.
\eeq
(However, the $\Rpv$ part of $\tilde{A}$ is
$ - \lambda^{i4i} \mu_i v_u/\sqrt{2} = m_{e_i} \mu_i \tan \beta$,
so appears positive in the amplitude...).

The Feynman rule we quote in figure 4
for the mass insertion $B_i$ also includes the
effect of the soft mass $m^2_{4i}$.  The minimisation
condition for the potential, in the $\snuvi = 0$ basis,
implies that
$m^2_{4i} = - B_i \tan \beta$ (at arbitrary
loop order) \cite{DLR}, so the  $\Rpv$ mixing
between $H^-$ and $E_L$ is
\beq
-B_i \cos \beta + m^2_{4i} \sin \beta = -\frac{B_i}{\cos \beta} ~~~.
\eeq

\begin{figure}[htb]
\unitlength1mm
\SetScale{2.8}
\begin{center}
\begin{picture}(80,20)(0,0)
\ArrowLine(0,0)(15,0)
\ArrowLine(30,0)(15,0)
\Text(-2,0)[r]{$\tilde{h}_d$}
\Text(32,0)[l]{$\tilde{h}_u$}
\Text(15,5)[c]{$\mu_i$}
\Text(15,0)[c]{x}
\Text(50,0)[l]{$ i \mu_i $} 
\end{picture}
\begin{picture}(80,20)(0,0)
\DashArrowLine(0,0)(15,0){1}
\DashArrowLine(30,0)(15,0){1}
\Text(-2,0)[r]{$H^+$}
\Text(32,0)[l]{$E_L$}
\Text(15,5)[c]{$B_i, m^2_{4i}$}
\Text(15,0)[c]{x}
\Text(50,0)[l]{$ i \frac{B_i}{ \cos \beta}$} 
\end{picture}
\begin{picture}(80,20)(0,0)
\DashArrowLine(0,0)(15,0){1}
\DashArrowLine(30,0)(15,0){1}
\Text(-2,0)[r]{$H^-$}
\Text(32,0)[l]{$E^c$}
\Text(15,5)[c]{$\tilde{A}$}
\Text(15,0)[c]{x}
\Text(50,0)[l]{$- i \sum_k \mu_k (m_{e_k} \tan \beta) \sin \beta $} 
\end{picture}
\end{center}
\vspace{1cm}
\label{intmass}
\caption{Internal charged line mass insertions.}
\end{figure}

\begin{figure}[htb]
\unitlength1mm
\SetScale{2.8}
\begin{center}
\begin{picture}(80,40)(0,10)
\ArrowLine(0,0)(15,0)
\ArrowLine(25,-10)(15,0)
\DashArrowLine(25,10)(15,0){1}
\Text(-2,0)[r]{$\nu_i$}
\Text(27,-10)[l]{$(e_L)_j$}
\Text(27,10)[l]{$E^c_k$}
\Text(50,0)[l]{$i \lambda^{ijn} E_{nk} $} 
\end{picture}
\begin{picture}(80,40)(0,10)
\ArrowLine(0,0)(15,0)
\ArrowLine(25,-10)(15,0)
\DashArrowLine(25,10)(15,0){1}
\Text(-2,0)[r]{$\nu_i$}
\Text(27,-10)[l]{$\tilde{h}_d^-$}
\Text(27,10)[l]{$E^c_k$}
\Text(50,0)[l]{$ i h_e^{i} E_{ik} $} 
\end{picture}
\begin{picture}(80,40)(0,10)
\ArrowLine(0,0)(15,0)
\ArrowLine(25,-10)(15,0)
\DashArrowLine(25,10)(15,0){1}
\Text(-2,0)[r]{$\nu_i$}
\Text(27,-10)[l]{$e^c_k$}
\Text(27,10)[l]{$(E_L)_m$}
\Text(50,0)[l]{$i \lambda^{ijk} L_{jm} $} 
\end{picture}
\begin{picture}(80,40)(0,10)
\ArrowLine(0,0)(15,0)
\ArrowLine(25,-10)(15,0)
\DashArrowLine(25,10)(15,0){1}
\Text(-2,0)[r]{$\nu_i$}
\Text(27,-10)[l]{$e^c_k$}
\Text(27,10)[l]{$H^-$}
\Text(50,0)[l]{$i h_e^{k}\sin \beta  $} 
\end{picture}
\end{center}
\vspace{2cm}
\label{feyn}
\caption{Feynman rules for trilinear/Yukawa interactions.
For the quarks, replace $\lambda_{ijk}
\rightarrow \lambda^{'ijk}$, $E^c_k \rightarrow D^c_k$,
$(e_L)_j \rightarrow (q_L)_j$, and so on.}
\end{figure}
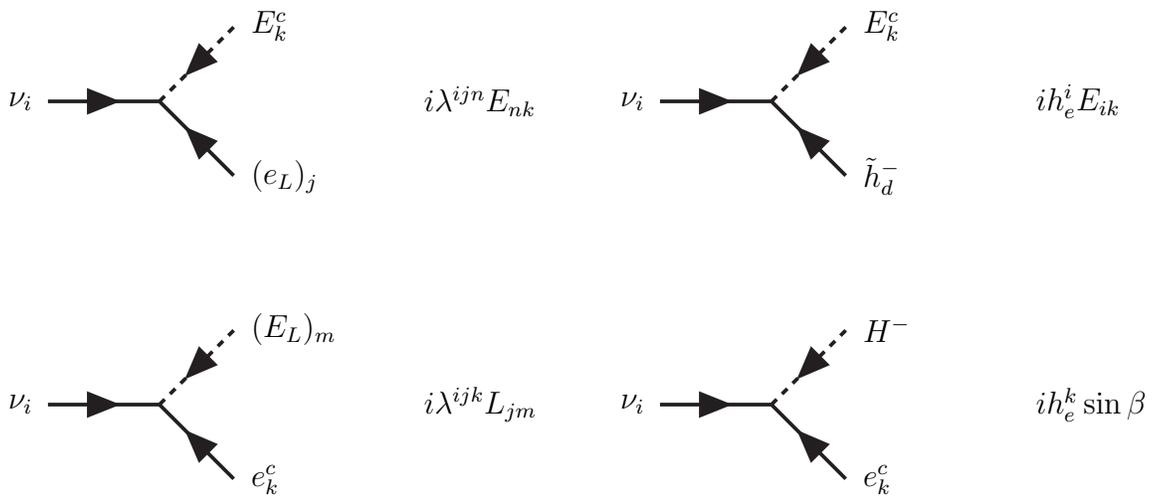

\section{Numerical Bounds}
\label{D1}

We give results for two different values of $\tan\beta =2, 10$ whenever necessary. This gives us an indication of the explicit dependence of the bounds
of $\tan\beta$, although there is another dependence implicit in the
neutralino/chargino mixing matrices which we do not address here.
Once again, we emphasize that we make the assumptions described in
section \ref{pheno} in order that
the combination of couplings constants are allowed the largest possible values.
It is easy to see that under these assumptions the integrals that appear
in the expressions of appendices A and B can be simply replaced by

\bea
I(m_1,m_2,m_3) &\rightarrow& {1\over 16\pi^2} {1\over m_{\mathrm susy}^{2}}, \nonumber \\
I(m_1,m_2,m_3,m_4) &\rightarrow& {1\over 16\pi^2} {1\over m_{\mathrm susy}^{4}}, \nonumber \\
I(m_1,m_2,m_3,m_4,m_5) &\rightarrow& {1\over 16\pi^2} {1\over m_{\mathrm susy}^{6}}. \nonumber \\
\eea
Below for all bounds we  use $|m_{\nu}|$.

\begin{itemize}

\item Tree-level contribution

\beq
|m_{\nu}^{ij}| = \delta_{\mu}^{i} \delta_{\mu}^{j }  m_{\mathrm susy},
\eeq

which gives the constraint

\beq
 \delta_{\mu}^{i} \delta_{\mu}^{j } \leq 10^{-12}.
\eeq

\item Diagram 1

\beq
m_{\nu}^{ij} = {1\over 8\pi^2} {1\over m_{\mathrm susy}}\sum_{k,n} \delta_{\lambda}^{ink}
 \delta_{\lambda}^{jkn} m_{e_{n}} m_{e_{k}}.
\eeq
Using the mass hierarchy of the lepton sector  for $i,j \neq 3$
we get

\bea
\delta_{\lambda}^{i33} \delta_{\lambda}^{j33}& \leq& 0.1 {\mathrm eV} 8\pi^2 { m_{\mathrm susy}\over m_{\tau}^2}, \nonumber \\
\delta_{\lambda}^{i33} \delta_{\lambda}^{j33}& \leq& 2.7 \times 10^{-7}, 
\eea
for $m_{\mathrm susy}= 100 $GeV.

For $i=1,2, j=3$ and $j=1,2, i=3$

\bea
\delta_{\lambda}^{i32} \delta_{\lambda}^{j23}& \leq& 0.1 {\mathrm eV} 8\pi^2 { m_{\mathrm susy}\over m_{\tau} m_{\mu}}, \nonumber \\
\delta_{\lambda}^{i33} \delta_{\lambda}^{j33}& \leq& 4.4 \times 10^{-6},
\eea
for $m_{\mathrm susy}= 100 $GeV.

For $i,j =3$

\bea
\delta_{\lambda}^{i22} \delta_{\lambda}^{j22}& \leq& 0.1 {\mathrm eV} 8\pi^2 { m_{\mathrm susy}\over  m_{\mu}^2}, \nonumber \\
\delta_{\lambda}^{i33} \delta_{\lambda}^{j33}& \leq& 7.1 \times 10^{-5}, 
\eea
for $m_{\mathrm susy}= 100 $GeV.
We can summarize the constraints in tables in the following way,

\beq
 \delta_{\lambda}^{ink}
 \delta_{\lambda}^{jkn}= \begin{array}{|l|c|c|c|} \hline
i/j & 1&2&3\\ \hline
1& 2.7 \times 10^{-7} & 2.7 \times 10^{-7} & 4.4 \times 10^{-6}\\ \hline
2 &  2.7 \times 10^{-7} & 2.7 \times 10^{-7} & 4.4 \times 10^{-6}\\ \hline
3& 4.4 \times 10^{-6}& 4.4 \times 10^{-6}& 7.1 \times 10^{-5} \\ \hline
\end{array}
\eeq

\item Diagram 2

\beq
m_{\nu}^{ij} = {3\over 8\pi^2} {1\over m_{\mathrm susy}}\sum_{k,n} \delta_{\lambda'}^{ink}
 \delta_{\lambda'}^{jkn} m_{d_{n}} m_{e_{k}}.
\eeq
Using the mass hierarchy of the down quark sector we get,

\bea
\delta_{\lambda'}^{i33} \delta_{\lambda'}^{j33}& \leq& 0.1 {\mathrm eV} 8\pi^2 { m_{\mathrm susy}\over 3 m_{b}^2}, \nonumber \\
\delta_{\lambda'}^{i33} \delta_{\lambda'}^{j33}& \leq& 1.05 \times 10^{-8}, 
\eea
for $m_{\mathrm susy}= 100 $GeV.

\item Diagram 3

\beq
m_{\nu}^{ij} = {g^2\over 64\pi^2 \cos^2\beta} m_{\mathrm susy} \delta_{B}^{i} \delta_{B}^{j}.
\eeq

This gives 

\bea
\delta_{B}^{i} \delta_{B}^{j} &\leq & 0.1 {\mathrm eV} {64 \pi^2 \cos^2\beta\over g^2 m_{\mathrm susy} }, \nonumber \\
\delta_{B}^{i} \delta_{B}^{j} &\leq & 2.9 \times 10^{-10},
\eea
for $\tan\beta=2$. 

\item Diagram 4

\beq 
m_{\nu}^{ij} = \sum_{k} {3 h_{d}^{k} m_{d_{k}}^{2}\over 16\pi^2  m_{\mathrm susy}} (\delta_{\mu}^{j}\delta_{\lambda'}^{ikk} + \delta_{\mu}^{i}\delta_{\lambda'}^{jkk}).
\eeq

Once again using the fermion mass hierarchy we get,

\beq 
m_{\nu}^{ij} = {3 h_{b} m_{b}^{2}\over 16\pi^2  m_{\mathrm susy}} (\delta_{\mu}^{j}\delta_{\lambda'}^{i33} + \delta_{\mu}^{i}\delta_{\lambda'}^{j33}).
\eeq
Which gives,

\beq
(\delta_{\mu}^{j}\delta_{\lambda'}^{i33} + \delta_{\mu}^{i}\delta_{\lambda'}^{j33}) \leq 
0.1 {\mathrm eV} {16\pi^2  m_{\mathrm susy}\over  3 h_{b} m_{b}^{2}}. 
\eeq

For $\tan\beta=2$,
\beq
(\delta_{\mu}^{j}\delta_{\lambda'}^{i33} + \delta_{\mu}^{i}\delta_{\lambda'}^{j33}) \leq 3.2 \times 10^{-7}.
\eeq

For $\tan\beta=10$,
\beq
(\delta_{\mu}^{j}\delta_{\lambda'}^{i33} + \delta_{\mu}^{i}\delta_{\lambda'}^{j33}) \leq 7.2 \times 10^{-8}.
\eeq

\item Diagram 5
\beq 
m_{\nu}^{ij} = {m_{\tau}\tan\beta\over 16\pi^2  m_{\mathrm susy}} \delta_{\lambda}^{ij3}\delta_{B}^{3}(m_{e_{j}} h_{e}^{j} -m_{e_{i}} h_{e}^{i}).
\eeq

For $\tan\beta =2$,

\beq
 \delta_{\lambda}^{ij3}\delta_{B}^{3} =\begin{array}{|l|c|c|c|} \hline
i/j & 1&2&3\\ \hline
1& 0 & 3.2 \times 10^{-3} & 1.2 \times 10^{-5}\\ \hline
2 &  3.2 \times 10^{-3} & 0 & 1.2 \times 10^{-5}\\ \hline
3& 1.2 \times 10^{-5}& 1.2 \times 10^{-5}& 0 \\ \hline
\end{array}
\eeq
For $\tan\beta =10$,

\beq
 \delta_{\lambda}^{ij3}\delta_{B}^{3} =\begin{array}{|l|c|c|c|} \hline
i/j & 1&2&3\\ \hline
1& 0 & 1.4 \times 10^{-4} & 5.5 \times 10^{-7}\\ \hline
2 &  1.4 \times 10^{-4} & 0 & 5.5 \times 10^{-7}\\ \hline
3& 5.5 \times 10^{-7}& 5.5 \times 10^{-7}& 0 \\ \hline
\end{array}
\eeq

\item Diagram 6

\beq
m_{\nu}^{ij} = \sum_{k} {h_{e}^{k} m_{e_{k}}^{2}\over 16\pi^2 m_{\mathrm susy}}( \delta_{\mu}^{j}
\delta_{\lambda}^{ikk} + \delta_{\mu}^{i}
\delta_{\lambda}^{jkk}).
\eeq

For $\tan\beta =2$,

\beq
( \delta_{\mu}^{j}
\delta_{\lambda}^{ikk} + \delta_{\mu}^{i}
\delta_{\lambda}^{jkk}) =\begin{array}{|l|c|c|c|} \hline
i/j & 1&2&3\\ \hline
1& 2.5 \times 10^{-5} & 2.5 \times 10^{-5} & 2.5 \times 10^{-5}\\ \hline
2 & 2.5 \times 10^{-5} & 2.5 \times 10^{-5} & 2.5 \times 10^{-5}\\ \hline
3& 2.5 \times 10^{-5}& 2.5 \times 10^{-5}& 1.0 \times 10^{-1} \\ \hline
\end{array}
\eeq

For $\tan\beta =10$,

\beq
( \delta_{\mu}^{j}
\delta_{\lambda}^{ikk} + \delta_{\mu}^{i}
\delta_{\lambda}^{jkk}) =\begin{array}{|l|c|c|c|} \hline
i/j & 1&2&3\\ \hline
1& 5.5 \times 10^{-6} & 5.5 \times 10^{-6} & 5.5 \times 10^{-2}\\ \hline
2 & 5.5 \times 10^{-6} & 5.5 \times 10^{-6} & 5.5 \times 10^{-2}\\ \hline
3& 5.5 \times 10^{-2}& 5.5 \times 10^{-2}& 2.3 \times 10^{-2} \\ \hline
\end{array}
\eeq

\item Diagram 7

\beq
m_{\nu}^{ij} = {\tan\beta \sin^2\beta\over 16\pi^2 m_{\mathrm susy} } \delta_{\mu}^{i}\delta_{\mu}^{j} (m_{e_{j}}^{2}
h_{e}^{j^{2}} + m_{e_{i}}^{2}
h_{e}^{i^{2}}).
 \eeq

For $\tan\beta =2$,

\beq
 \delta_{\mu}^{i}\delta_{\mu}^{j}  =\begin{array}{|l|c|c|c|} \hline
i/j & 1&2&3\\ \hline
1& --- & --- & 7.1 \times 10^{-4}\\ \hline
2 & --- &--- & 7.1 \times 10^{-4}\\ \hline
3& 7.1 \times 10^{-4}& 7.1 \times 10^{-4}& 3.5 \times 10^{-4} \\ \hline
\end{array}
\eeq

For $\tan\beta =10$,

\beq
 \delta_{\mu}^{i}\delta_{\mu}^{j}  =\begin{array}{|l|c|c|c|} \hline
i/j & 1&2&3\\ \hline
1& --- & --- & 5.7 \times 10^{-6}\\ \hline
2 & --- & 0.35 & 5.7 \times 10^{-6}\\ \hline
3& 5.7 \times 10^{-6}& 5.7 \times 10^{-6}& 2.8 \times 10^{-6} \\ \hline
\end{array}
\eeq

The dashed lines mean that there is no bound.

\item Diagram 8

\beq 
m_{\nu}^{ij} = \sum_{k} \delta_{\mu}^{k} m_{e_{k}} {\tan\beta \sin^2\beta\over 16\pi^2 m_{\mathrm susy}  } \delta_{\lambda}^{ijk} (m_{e_{j}} h_{e}^{j} - m_{e_{i}} h_{e}^{i}).
\eeq

The bounds can be obtained from the bounds of diagram 5$/(\sin^2\beta)$.

\item Diagram 9

\beq 
m_{\nu}^{ij} = {\tan\beta \sin^2\beta \over 8\pi^2  m_{\mathrm susy} } \delta_{\mu}^{i}\delta_{\mu}^{j} m_{e_{i}} m_{e_{j}} h_{e}^{i} h_{e}^{j}.
\eeq

For $\tan\beta=2$,

\beq 
 \delta_{\mu}^{i}\delta_{\mu}^{j} =\begin{array}{|l|c|c|c|} \hline
i/j & 1&2&3\\ \hline
1& --- & --- & ---\\ \hline
2 & --- &--- & 9.3 \times 10^{-2}\\ \hline
3& --- & 9.3 \times 10^{-2}& 3.5 \times 10^{-4} \\ \hline
\end{array}
\eeq

For $\tan\beta=10$,

\beq 
 \delta_{\mu}^{i}\delta_{\mu}^{j} =\begin{array}{|l|c|c|c|} \hline
i/j & 1&2&3\\ \hline
1& --- & --- & ---\\ \hline
2 & --- &0.19 & 7.4 \times 10^{-4}\\ \hline
3& ---& 7.4 \times 10^{-4}& 2.8 \times 10^{-6} \\ \hline
\end{array}
\eeq

\item Diagram 10

\beq 
m_{\nu}^{ij} = {1 \over 16\pi^2  m_{\mathrm susy} } \delta_{\mu}^{i}\delta_{\mu}^{j} ( m_{e_{i}}^{2} h_{e}^{i^{2}} +  m_{e_{j}}^{2} h_{e}^{j^{2}}).
\eeq

The bounds here can be obtained from those diagram 7$\times \tan\beta \sin^2\beta$.

\item Diagram 11

\beq
m_{\nu}^{ij} = {\tan\beta \over 16\pi^2  m_{\mathrm susy} } \delta_{\mu}^{i}\delta_{B}^{j} ( m_{e_{j}}^{2} h_{e}^{j^{2}}) +  \delta_{\mu}^{j}\delta_{B}^{i}( m_{e_{i}}^{2} h_{e}^{i^{2}})
\eeq

The bounds here can be obtained from those diagram 7$\times  \sin^2\beta$.

\item Diagram 12

\beq
m_{\nu}^{ij} = \sum_{k} {m_{e_{k}}\over 16\pi^2  m_{\mathrm susy} } \delta_{\mu}^{k}\delta_{\lambda}^{ijk} (m_{e_{i}} h_{e}^{i} - 
m_{e_{j}} h_{e}^{j}).
\eeq

The bounds here can be obtained from those diagram 5$\times \tan\beta$.

\item Diagram 13

\beq 
m_{\nu}^{ij} = {1 \over 8\pi^2  m_{\mathrm susy} } \delta_{\mu}^{i}\delta_{B}^{j} ( m_{e_{i}} h_{e}^{i} m_{e_{j}} h_{e}^{j}).
\eeq

The bounds here can be obtained from those diagram 9$\times \tan\beta \sin^2\beta$.

\item Diagram 14

\beq 
m_{\nu}^{ij} = {g \tan\beta \over 16\pi^2 \sqrt 2 }( \delta_{\mu}^{i}\delta_{B}^{j}  m_{e_{i}} h_{e}^{i} + \delta_{\mu}^{j }\delta_{B}^{i} m_{e_{j}} h_{e}^{j}).
\eeq

For $\tan\beta=2$,

\beq 
( \delta_{\mu}^{i}\delta_{B}^{j})
=\begin{array}{|l|c|c|c|} \hline
i/j & 1&2&3\\ \hline
1& --- & 1.2 \times 10^{-4} & 4.6 \times 10^{-7}\\ \hline
2 & --- &6.0 \times 10^{-5} & 4.6 \times 10^{-7}\\ \hline
3& --- & --- & 2.3 \times 10^{-7} \\ \hline
\end{array}
\eeq
The $i$ index on $\delta_{\mu}$ corresponds to the column, and
the $j$ index to the row, for both the table
above and below.

For $\tan\beta=10$,

\beq 
( \delta_{\mu}^{i}\delta_{B}^{j} )   =\begin{array}{|l|c|c|c|} \hline
i/j & 1&2&3\\ \hline
1& 0.12 & 5.3 \times 10^{-6} & 2.0 \times 10^{-8}\\ \hline
2 & --- &2.6 \times 10^{-6} & 2.0 \times 10^{-8}\\ \hline
3& --- & --- & 1.0 \times 10^{-8} \\ \hline
\end{array}
\eeq

\item Diagram 15 

\beq 
m_{\nu}^{ij} = {g\over 16\pi^2 \sqrt 2} \delta_{\mu}^{i} \delta_{\mu}^{j} (  m_{e_{i}} h_{e}^{i} + 
 m_{e_{j}} h_{e}^{j}).
\eeq

The bounds here can be obtained from those diagram 20$\times \tan\beta \sin^2\beta$.

\item Diagram 16

\beq 
m_{\nu}^{ij} = {g \sin^2 \beta \over 16\pi^2 \sqrt 2} \delta_{\mu}^{i} \delta_{\mu}^{j} (  m_{e_{i}} h_{e}^{i} + 
 m_{e_{j}} h_{e}^{j}).
\eeq

The bounds here can be obtained from those diagram 20$\times \tan\beta$.

\item Diagram 17

From the slepton-lepton loop, neglecting
$g'$ with respect to $g$:

\beq
m_{\nu}^{ij} = \sum_{k} {g m_{e_{k}}\over 16\pi^2 \sqrt 2} 
 (\delta_{\mu}^{i}\delta_{\lambda}^{jkk} 
+ \delta_{\mu}^{j}\delta_{\lambda}^{ikk})
\eeq

\beq
 (\delta_{\mu}^{i} \delta_{\lambda}^{j33} + \delta_{\mu}^{j} \delta_{\lambda}^{i33}) = 1.0 \times 10^{-8}
\eeq

From the squark-quark loop:

\beq
m_{\nu}^{ij} = 3 \sum_{k} {g m_{d_{k}}\over 16\pi^2 \sqrt 2} 
 (\delta_{\mu}^{i}\delta_{\lambda}^{'jkk} 
+ \delta_{\mu}^{j}\delta_{\lambda}^{'ikk})
\eeq

\beq
 (\delta_{\mu}^{i} \delta_{\lambda}^{'j33} 
+ \delta_{\mu}^{j} \delta_{\lambda}^{'i33}) = 3.4 \times 10^{-9}
\eeq

\item Diagram 18

This diagram is non-zero only for non-degenerate sleptons.

\item Diagram 19

\beq
m_{\nu}^{ij} = {g^2\over 64\pi^2 \cos\beta} m_{\mathrm susy} (\delta_{\mu}^{i} \delta_{B}^{j} +
\delta_{\mu}^{j} \delta_{B}^{i}). 
\eeq

For $\tan\beta =2$,

\beq
 (\delta_{\mu}^{i} \delta_{B}^{j} +
\delta_{\mu}^{j} \delta_{B}^{i}) = 6.7 \times 10^{-10}
\eeq

For $\tan\beta =10$,

\beq
 (\delta_{\mu}^{i} \delta_{B}^{j} +
\delta_{\mu}^{j} \delta_{B}^{i}) = 1.4 \times 10^{-10}
\eeq

\item Diagram 20

\beq 
m_{\nu}^{ij} = {g \tan\beta \sin^2\beta \over 16\pi^2 \sqrt 2} \delta_{\mu}^{i} \delta_{\mu}^{j}(
m_{e_{i}} h_{e}^{i} + m_{e_{j}} h_{e}^{j} ).
\eeq

For $\tan\beta=2$,

\beq
 \delta_{\mu}^{i} \delta_{\mu}^{j} =\begin{array}{|l|c|c|c|} \hline
i/j & 1&2&3\\ \hline
1& --- & 1.5 \times 10^{-4} & 5.7 \times 10^{-7}\\ \hline
2 & 1.5 \times 10^{-4} & 7.5 \times 10^{-5} & 5.7 \times 10^{-7}\\ \hline
3& 5.7 \times 10^{-7}& 5.7 \times 10^{-7}& 2.8 \times 10^{-7} \\ \hline
\end{array}
\eeq

For $\tan\beta=10 $,

\beq
 \delta_{\mu}^{i} \delta_{\mu}^{j} =\begin{array}{|l|c|c|c|} \hline
i/j & 1&2&3\\ \hline
1& 0.12 & 5.4 \times 10^{-6} & 2.0 \times 10^{-8}\\ \hline
2 & 5.4 \times 10^{-6} & 2.7 \times 10^{-6} & 2.0 \times 10^{-8}\\ \hline
3& 2.0 \times 10^{-8}& 2.0 \times 10^{-8}& 1.0 \times 10^{-8} \\ \hline
\end{array}
\eeq

\item Diagram 21

\beq 
m_{\nu}^{ij} = {g \tan\beta \over 16\pi^2 \sqrt 2} (\delta_{\mu}^{i} \delta_{B}^{j}
m_{e_{j}} h_{e}^{j} + \delta_{\mu}^{j} \delta_{B}^{i} m_{e_{i}} h_{e}^{i} ).
\eeq

Here we obtain the same bounds as in diagram 14, but with
the indices on $\delta_{\mu}$ and  $\delta_{B}$ inverted;
the bound of diagram 14  on $\delta_{\mu}^i\delta_{B}^j$
applies to  $\delta_{\mu}^j\delta_{B}^i$ according
to this diagram.

\end{itemize}

\end{document}